\documentclass[useAMS,referee]{biom}
\usepackage{amsmath}
\usepackage{amssymb}
\usepackage{booktabs}
\usepackage{algorithm}  
\usepackage{diagbox}
\usepackage{url}
\usepackage{algpseudocode}  
\usepackage{apalike}
\let\bibhang\relax
\usepackage{natbib}
\usepackage{threeparttable}
\usepackage{graphicx}
\usepackage[table,xcdraw]{xcolor}
\def\bSig\mathbf{\Sigma}

\newtheorem{Assumption}{Assumption}
\newtheorem{Lemma}{Lemma}[section]
\newtheorem{Theorem}{Theorem}[section]

\title{Causal inference with outcome dependent sampling and mismeasured outcome}

\author{Min Zeng$^{1}$, 
Zeyang Jia$^{1}$, 
Zijian Sui$^{1}$, 
Jinfeng Xu$^{2,*}$\email{jinfenxu@cityu.edu.hk}, and 
Hong Zhang$^{1,**}$\email{zhangh@ustc.edu.cn} \\
$^{1}$Department of Statistics and Finance, University of Science and Technology of China, Anhui, China\\
$^{2}$Department of Biostatistics, City University of Hong Kong, Hong Kong, China.}

\begin{document}

\date{{\it Received October} 2007. {\it Revised February} 2008.  {\it
Accepted March} 2008.}



\pagerange{\pageref{firstpage}--\pageref{lastpage}} 
\volume{64}
\pubyear{2008}
\artmonth{December}


\doi{10.1111/j.1541-0420.2005.00454.x}


\label{firstpage}


\begin{abstract}
Outcome-dependent sampling designs are extensively utilized in various scientific disciplines, including epidemiology, ecology, and economics, with retrospective case-control studies being specific examples of such designs. Additionally, if the outcome used for sample selection is also mismeasured, then it is even more challenging to estimate the average treatment effect (ATE) accurately. To our knowledge, no existing method can address these two issues simultaneously. In this paper, we establish the identifiability of ATE and propose a novel method for estimating ATE in the context of generalized linear model. The estimator is shown to be consistent under some regularity conditions. To relax the model assumption, we also consider generalized additive model. We propose to estimate ATE using penalized B-splines and establish asymptotic properties for the proposed estimator. Our methods are evaluated through extensive simulation studies and the application to a dataset from the UK Biobank, with alcohol intake as the treatment and gout as the outcome.
\end{abstract}

%

\begin{keywords}
 Average treatment effect; Causal inference; Outcome dependent sampling; Outcome mismeasurement.
\end{keywords}


\maketitle


%
\section{Introduction}
\label{s:intro}

Numerous studies in the fields of biomedical and social sciences are focused on determining the causal impact of a binary treatment on a specific outcome. Although randomized controlled trials (RCTs) serve as the gold standard for establishing causal relationships, they may not always be feasible due to financial, logistical, or ethical constraints. As a result, researchers often rely on observational studies. Various methodologies, such as propensity score techniques \citep{a1, a2} and instrumental variable estimation methods \citep{a3, a4}, have been developed to estimate average treatment effect (ATE) in observational studies. However, the efficacy of these methodologies relies on sampling randomness. In cases where sample selection is not random, these methods are no longer valid.

Outcome-dependent sampling (ODS) represents a non-random sampling design in which the selection of sample units depends on the outcome of interest. ODS offers some advantages over simple random sampling, such as enhanced statistical power in the situation where the outcome is rare \citep{b1}. However, ODS greatly complicates the statistical data analysis and result interpretation. 
The case-control design, along with its variations, is the most prevalent form of ODS design. Ideally, in such designs, the sampling process relies exclusively on the outcome rather than any other variables in the study. If the unique characteristics of ODS designs are not considered, the conventional causal inference methods may be subject to selection bias \citep{a5,a6}.
A large quantity of research based on ODS designs has been published\citep{a7, a8}, but the majority of these studies focus on association analysis instead of causal inference.
Several researchers \citep{a12,a10,a9} have attempted to avoid this issue by focusing on causal risk ratio. \cite{a13} and \cite{b2}, on the other hand, proposed an ATE estimator based on a weighted targeted maximum likelihood by incorporating information on disease prevalence.

However, implementing an ideal ODS design in practice can often be challenging, as sample selection may be influenced, at least partially, by diagnosis or measurement. As a result, the true outcome of interest may be unobserved, and the measured outcome may differ from the true outcome. 
Various factors contribute to mismeasurement in outcome variables, such as the unavailability of costly measurements, the propensity to misreport responses to sensitive questions, and the inevitable recall bias. 
Numerous studies have investigated the impact of mismeasurement of outcome variables such as bias and efficiency loss \citep{a18,a19,a20,b3}. Some researchers have opted to develop sensitivity analyses of mismeasured binary outcomes to reduce bias \citep{a14,a15,a16}. \cite{a20, a21, a22} derived asymptotic bias of the conventional inverse probability of treatment weighting (IPTW) and doubly robust (DR) estimators ignoring the measurement error, and proposed a modified weighting method by rectifying the mismeasurement bias.

Although research on addressing either selection bias or mismeasurement bias has garnered much attention, very few methods have been developed to deal with these two types of bias simultaneously except for \cite{a28} and \cite{a24}. However, these studies focus on association analysis. 
To our knowledge, no causal inference method has been developed to simultaneously address both issues.
 In this paper, we derived a novel generalized linear model (GLM) to establish the relationship between the observed samples and the target population. This allows for an intuitive understanding of the combined effects of ODS and measurement error on ATE estimation.
Then we derive estimation equations (EE) to estimate unknown parameters, through which we obtain an ATE estimator. We call this method GLM-EE. The GLM-EE estimator is proven to be consistent and asymptotically normal. Furthermore, to relax model assumption, we introduce a generalized additive model (GAM) based estimator \citep{a51, a31, a32}, which employs penalized spline as a smoothing technique. 
Unknown parameters can again be estimated by solving a set of estimation equations and we refer to this method as GAM-EE. Asymptotic properties of the GAM-EE estimator are also established. 
Through simulation study, we demonstrate that both proposed estimators effectively address selection bias and mismeasurement bias. Moreover, the GAM-EE method is shown to be more robust to model misspecification with little efficiency sacrifice.

We further applied our method to a real-world dataset from the UK Biobank, which aims to investigate the ATE of alcohol consumption (treatment) on gout disease (outcome) among male individuals aged 40 to 80. Gout is a form of arthritis that arises when uric acid crystals accumulate in the joints, causing inflammation and pain. However, this outcome measurement suffers from misdiagnoses with a high false negative rate of 10-30\% and a low false positive rate of about $5\%$ \citep{a34, a33}. We applied our methods to this dataset and conducted a sensitivity analysis to evaluate their performance in real-world research.

The remainder of this article is structured as follows. In Section 2, we introduce our model and establish the identifiability of ATE under some appropriate assumptions. In Section 3, we describe our GLM-EE  and GAM-EE methods and establish their theoretical properties. In Section 4 and Section 5, we evaluate the performance of the two methods through extensive simulation studies and a real data application, respectively. In Section 6, we conclude the article and discuss future research directions. All technical details, along with supplementary information for the numerical studies in Sections 4 and 5, are provided in Supplementary Materials.

\section{Identifiability of average treatment effect}
\label{s:identification}

\subsection{Ordinary scenario}

We start by reviewing the identifiability of ATE in ordinary scenarios where samples are selected randomly and measurement error is absent.  
Let $T$ and $Y$ denote the binary treatment and true outcome of interest, respectively. Let $Y(t)$ denote the potential or counterfactual outcome for a given subject with exposure level $t$ \citep{a30}. Suppose $Y$, $T$, and $Y(t)$ take values in the binary set $\{0,1\}$. Let $X$ denote a vector of covariates or confounding variables. Our target parameter ATE, denoted as $\tau$, is defined as the expected difference between the potential outcomes:
 $$
\tau = \mathbb E[Y(1)-Y(0)],
 $$
where the expectation is evaluated in the target population. 
 
In the standard causal inference framework, the identifiability of $\tau$ hinges on three fundamental assumptions stated as follows:
\begin{Assumption}
(consistency)
\label{asm: 1}
$$
Y = TY(1) + (1-T)Y(0).
$$
\end{Assumption}
\begin{Assumption}
(positivity)
\label{asm: 2}
$$
1>\mathbb P(T=1|X)>0.
$$
\end{Assumption}
\begin{Assumption}
(unconfoundness)
\label{asm: 3}
$$
(Y(1), Y(0)) \perp T \mid X.
$$
\end{Assumption}
Under Assumptions \ref{asm: 1}-\ref{asm: 3}, $\tau$ is identifiable, as demonstrated by the following formula:
\begin{align}
\tau = \mathbb E\{g_1(X) - g_0(X)\} \label{eq:tau_definition},
\end{align}
where $g_i(x) = \mathbb E[Y \mid X=x, T=i], \ i=1, 0$.

As evident from Equation (\ref{eq:tau_definition}), the identifiability of $\tau$ relies not only on $g_1$ and $g_0$ but also on the distribution of covariates $X$ in the target population. 
In scenarios involving ODS designs and measurement error, both $g_i(x)$ and the distribution of $X$ are ambiguous. Consequently, the identifiability of $\tau$ needs further assumptions.

\subsection{ODS with measurement error}

\begin{figure}
\centering
\includegraphics[width=13cm, height=7cm]{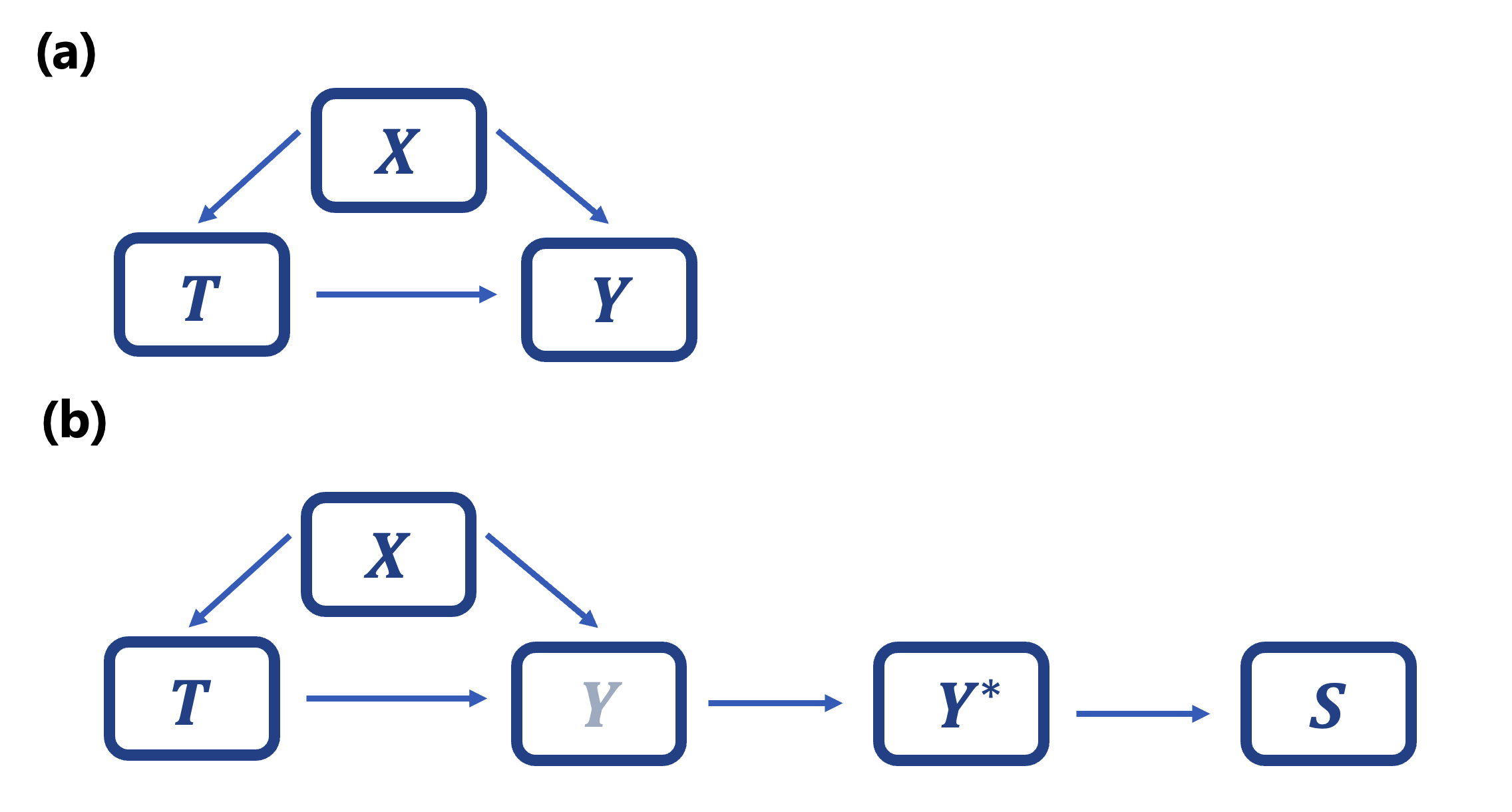}
\caption{Causal directed acyclic graph (a) is the usual causal inference framework, while (b) is the ODS design with the mismeasured outcomes. The light color indicates the latent variable, and the dark color indicates the observed variables.}
\label{f:figure1}
\end{figure}

Let  $Y^*$ denote the observed outcome, which may differ from the true outcome $Y$ due to measurement error. Let $S$ represent an indicator of selection into the study, with $S=1$ for ``selected" and $S=0$ for ``not selected". The sample distribution is expressed by $\mathbb P(Y^*, X, T \mid S=1)$, where $\mathbb P(\cdot)$ denotes the probability density function.
To ensure the identifiability of $\tau$, we introduce two additional assumptions characterizing the mechanism of sample selection and outcome measurement. 
\begin{Assumption} (selection conditional independence): the sample selection procedure is independent of $(Y,X,T)$ given $Y^*$, that is
\label{asm: 4}
$$
S \perp (Y, X, T) \mid Y^*.
$$
\end{Assumption}
\begin{Assumption} (measurement conditional independence): the observed outcome $Y^*$ is independent of $(X,T)$ given $Y$, that is
\label{asm: 5}
$$
Y^* \perp (X, T) \mid Y.
$$
\end{Assumption}
Assumption \ref{asm: 4} naturally aligns with outcome-dependent sampling (ODS) design, as it posits that samples are selected solely based on the observed outcome $Y^*$. Assumption \ref{asm: 5} states that the observed outcome $Y^*$ relies exclusively on the true outcome $Y$, which indicates the influence of $X$ and $T$ on $Y^*$ is completely mediated by $Y$. 
This frequently arises in clinical diagnosis and is extensively employed in the literature \citep{a20, a21}. Figure \ref{f:figure1} employs the directed acyclic graph (DAG) to illustrate the problem, where subfigure (a) corresponds to the ordinary design under Assumptions \ref{asm: 1}-\ref{asm: 3}, while subfigure (b) depicts the ODS design with measurement error under Assumptions \ref{asm: 1}-\ref{asm: 5}.

It follows from Assumptions \ref{asm: 4} and \ref{asm: 5} that $\mathbb P(Y^* = j \mid Y=i, X) = \mathbb P(Y^* = j \mid Y=i)$ and $\mathbb P(S=j \mid Y^* = i, X)=\mathbb P(S=j \mid Y^* = i)$ for $i,j = 1$ or $0$. To simplify statement, we denote $p_{ij} = \mathbb P(Y^*=j|Y=i)$ for $i,j = 1$ or $0$, where $p_{01}$ is the false positive rate of the disease and $p_{10}$ is the false negative rate of the disease. 
Let $v = \mathbb P(Y=1)$ denote the disease prevalence in the target population. 
Since $v$, $p_{01}$, and $p_{10}$ are usually attainable through existing literature and medical expert consultations, we assume these values are known.

Let $s = {\mathbb{P}(S=1|Y^*=0)}/{\mathbb{P}(S=1|Y^*=1)}$ denote the sampling ratio between cases and controls. This ratio measures the degree of sampling bias, with $s=1$ indicating random sampling, and as $s$ deviates further from 1, the level of selection bias increases. Let $v^* = \mathbb P(Y^* = 1)$ denote the observed disease prevalence,  which may differ from $v$ due to measurement error.  
Let $g_i^*(x) = \mathbb{E}[Y^* | X=x, T=i, S=1]$ denote the expectation of $Y^*$ conditional on $X=x$, $T=t$, and $S=1$, which can be identified by the sample distribution. The following lemma explores the relationship between $g_i^*(x)$ and $g_i(x)$.
\begin{Lemma}
\label{Le:1}
Under Assumptions 1-5, for $i$ = $0$ or $1$, we have 
\begin{align}
\label{eq:identi1}
g_i^*(X)=  \frac{((1-p_{10}-p_{01})g_i(X)+p_{01})s}{1+((1-p_{10}-p_{01})g_i(X)+p_{01})(s-1)},
\end{align}
where 
\begin{align}
\label{eq: s}
s=\frac{\mathbb P(Y^*=1|S=1)/v^*}{\mathbb P(Y^*=0|S=1)/(1-v^*)},
\end{align}
\begin{align}
\label{eq: vstar}
v^*=(1-p_{10}-p_{01})v+p_{01}.
\end{align}
\end{Lemma}
Lemma \ref{Le:1} indicates that the sampling ratio $s$ and observed disease prevalence $v^*$ are determined by $v$, $p_{01}$ and $p_{10}$.
Also, there is a one-to-one function relationship between $g_i^*(X)$ and $g_i(X)$ given $v$, $p_{01}$ and $p_{10}$, which demonstrates that $g_i(X)$ is identifiable since $g_i^*(X)$ is determined by the sample distribution. 
To ensure the identifiability of $\tau$, one must also calculate the expectation of $g_i(X)$.

\begin{Lemma}
\label{Le:2}
Under Assumptions 1-5, for $i$ = $0$ or $1$, we have
\begin{align}
\mathbb E[g_i(X)]=v^*u_{i1}+(1-v^*)u_{i0},\label{eq:identi3}
\end{align}
where 
\begin{align}
    \label{eq:u}
    u_{ij} = \int g_i(x)f(x|Y^*=j,S=1)dx,
\end{align}
$v^*$ is given in (\ref{eq: vstar}) and $f(\cdot \mid Y^*, S)$ represents the conditional density of $X$ given $Y^*$ and $S$.
\end{Lemma}
The proof of Lemma \ref{Le:2} is straightforward by applying the law of total probability and Assumption \ref{asm: 4}. Applying Lemmas \ref{Le:1}-\ref{Le:2}, we can establish the identifiability of $\tau$, as described in Theorem \ref{th: identifiable}.
\begin{Theorem}
\label{th: identifiable}
Under Assumptions \ref{asm: 1}-\ref{asm: 5}, the average treatment effect $\tau$ is identifiable: 
$$\tau = \mathbb E[g_1(X)]- \mathbb E[g_0(X)],$$
where 
\begin{align}
\mathbb E[g_i(X)]&=\frac{v^*}{1-p_{10}-p_{01}}\int \left(\frac{g_i^*(x)}{s-g_i^*(x)(s-1)}-p_{01}\right)f(x|Y^*=j,S=1)dx \nonumber \\
&+\frac{1-v^*}{1-p_{10}-p_{01}}\int \left(\frac{g_i^*(x)}{s-g_i^*(x)(s-1)}-p_{01}\right)f(x|Y^*=j,S=1) dx,\  i = 1,\  0,
\label{eq:identi4}
\end{align}
where $s$ and $v^*$ are given in (\ref{eq: s}) and (\ref{eq: vstar}), respectively.
\end{Theorem}
Since $\mathbb P(Y^*, X, T|S=1)$ can be approximated by its sample version, we can consistently estimate ATE $\tau$ if $p_{01},p_{10}, v$ are given. We provide methods for estimating $\tau$ in the next section.

\section{Estimation of ATE $\tau$}
\label{s:model}

In this section, we derive the estimation bias of naive method ignoring selection sampling and mismeasurement, then propose two debias methods. Both methods depend on an adjusted link function associated with sampling ratio $s$ and mismeasurement probabilities $p_{01}$ and $p_{10}$. 

According to Theorem \ref{th: identifiable}, to estimate $\tau$, we can first estimate $g_i$ then estimate $\mathbb E[g_i(X)]$, $i=0,\ 1$. To begin with, we model $g_i$ with a logistic link:
$$
g_i(x) = \frac{\exp(\eta(T=i,X=x))}{1+\exp(\eta(T=i,X=x))},
$$
where the index $\eta(t, x)$ is a function of $t$ and $x$. Applying Lemma 2.1, we obtain that
$$ 
g_i^*(x)= h(\eta(T=i,X=x)),
$$ where
\begin{align}
\label{eq: link}
    h(\eta)=\frac{\left(p_{01}+\exp(\eta) (1-p_{10})\right)s}{1+p_{01}(s-1)+\exp(\eta)\left(1+(1-p_{10})(s-1)\right)}.
\end{align}
serves as an adjusted link function.The log-likelihood function of the observed samples is 
\begin{align}
\label{eq:likelihood}
     \ell_n\left(y^*, \eta(t, \boldsymbol x)\right) = \sum_{i = 1}^n \left\{y^*_i \log(\frac{\mu_i}{1-\mu_i}) + \log(1-\mu_i)\right\}, 
\end{align}
where $\mu_i = h(\eta(t_i, \boldsymbol x_i))$ for $i$-th sample and $n$ is the size of sample.

\textbf{Remark 1: }The adjusted link function $h(\eta)$ is a monotone increasing function of index $\eta$. The shape of its curve is highly influenced by the values of $p_{01}$, $p_{10}$, and $s$. For example, the function has a supremum of $\frac{(1-p_{10})s}{1+(1-p_{10})(s-1)}$ and an infimum of $\frac{p_{01}s}{1+p_{01}(s-1)}$. Figure \ref{f:figure2} in Appendix B of Supplementary Materials illustrates the curves of $h(\eta)$ for various combinations of $p_{01}$, $p_{01}$, and $s$. It is worth noting that when $p_{01}=0$, $p_{10}=0$, and $s=1$, $h(\eta)$ degenerate to the logistic link function, corresponding to the situation of random sampling and no measurement error. 

\textbf{Remark 2: } As shown in Figure \ref{f:figure2}, when $p_{01}>0$, $p_{10}>0$, and $s>1$, the shape of the curves becomes deflated and shifted. Especially when $s$ is large and $p_{01}$ is bigger than 0, the infimum of $h(\eta)$ approaches $1$. This leads to a significantly reduced range of the adjusted link function, which in turn makes model estimation challenging. To avoid an excessively small range of $h(\eta)$ when $s$ is big, it is crucial to ensure that $p_{01}$ does not deviate too far from 0.  Fortunately, in real world, $p_{01}$ is often very close to 0, maintaining a reasonable range for the $\eta$ function.

\textbf{Remark 3: }Consider the ODS design with no measurement error, that is 
$p_{01}=0$, $p_{10}=0$ and $s>1$. If we further assume that $\eta$ has a linear form $\eta(t,\boldsymbol x) = a_0 + a_1 T + \boldsymbol b^\mathsf{T} \boldsymbol x$, then the logistic regression produces consistent estimates of the slope parameters $a_1$ and $\boldsymbol{b}^\mathsf T$. However, when it comes to ODS design with measurement error, that is $p_{01}>0$, $p_{10}>0$ and $s>1$, all the logistic regression estimators will be biased. Details can be seen in \cite{a52}. 


\subsection{A generalized linear model based estimator }
\label{sub:GLM}

In this subsection, we assume that the true model for $\eta(t,\boldsymbol x)$ has a linear form: $\eta(t,\boldsymbol x) = a_0 + a_1 T + \boldsymbol b^\mathsf{T} \boldsymbol x$. Denote $\boldsymbol \beta^\mathsf{T} = (a_0, a_1, \boldsymbol b^\mathsf{T})$. The estimation of $\boldsymbol \beta$ can be performed by maximizing the log-likelihood function of GLM with the adjusted link function $h(\cdot)$. The estimation of $\tau$ can then be obtained by following the subsequent steps.

\textbf{Step 0}: \emph{Determine the values of $p_{10},p_{01}, v$.} 

\textbf{Step 1}: \emph{Estimate sampling ratio $s$ and the adjusted link function $h$.} 
Motivated by (\ref{eq: s}), we obtain an estimator of $s$ by solving the estimating equation.
\begin{align}
    {S_n}(s) := \frac{1}{n} \sum_{i = 1}^n\{s v^*    (1 - y_i^*) - (1 - v^*) y_i^* \} = 0\label{eq: sn1}.
\end{align}
Denote the resulting estimator by $\hat{s} = \frac{\sum_{i=1}^n y_i^*(1-v^*)}{\sum_{i=1}^n (1-y_i^*)v^*}$. Plugging $\hat{s}$ in (\ref{eq: link}), we obtain an estimator of the link function $h(\cdot)$, denoted as $\hat h(\cdot)$.

\textbf{Step 2}: \emph{Estimate $\boldsymbol \beta$.}
We can estimate $\boldsymbol \beta$ by solving the following score equations:
\begin{align}
   {G_n}(\boldsymbol \beta) :=\frac{1}{n} \frac{\partial \ell_n\left(\boldsymbol \beta\right)}{\partial \boldsymbol{\beta}}=0
   \label{eq: gn1},
\end{align}
where $l_n(\beta)$ is defined in (\ref{eq:likelihood}), with $h(\cdot)$ beging replaced by $\hat h(\cdot)$.
Denote the resulting estimator by  $\hat{\boldsymbol\beta}^\mathsf{T} = ( \hat a_0, \hat a_1, \hat {\boldsymbol b}^\mathsf{T})$.
Denote $\hat g_1(x) = \operatorname{expit}(\hat a_0 + \hat a_1  + \hat{\boldsymbol b}^\mathsf{T} x)$ and $\hat g_0(x) = \operatorname{expit}(\hat a_0 + \hat{\boldsymbol b}^\mathsf{T} x)$.

\textbf{Step 3}: \emph{Estimate $\boldsymbol u = (u_{11}, u_{10}, u_{01}, u_{00})^\mathsf{T}$.}
Inspired by (\ref{eq:u}), we can get the estimators of $u_{ij}$ by
$$
\hat{u}_{11}=\frac{1}{\sum_{i=1}^n y^*_i}\sum_{i=1}^{n}y_i^*\hat{g}_1(x_i),\ \ 
\hat{u}_{10}=\frac{1}{\sum_{i=1}^n (1-y^*_i)}\sum_{i=1}^{n}(1-y_i^*)\hat{g}_1(x_i),
$$ 
$$
\hat{u}_{01}=\frac{1}{\sum_{i=1}^n y^*_i}\sum_{i=1}^{n}y_i^*\hat{g}_0(x_i),\ \ 
\hat{u}_{00}=\frac{1}{\sum_{i=1}^n (1-y^*_i)}\sum_{i=1}^{n}(1-y_i^*)\hat{g}_0(x_i).
$$ 

\textbf{Step 4}: \emph{Estimate the average treatment effect $\tau$.}
According to (\ref{eq:identi3}), we can estimate $\tau$ by
$$
\hat{\tau}=\left[v^*\hat{u}_{11}+(1-v^*)\hat{u}_{10}\right]-\left[v^*\hat{u}_{01}+(1-v^*)\hat{u}_{00}\right],
$$
where $v^*$ is calculated according to (\ref{eq: vstar}).

To establish the asymptotic distribution of estimated parameters, we write steps 1-3 in a form of estimation equations:
\begin{align}
\label{eq:psi1}
\left(\begin{array}{c}
{S_n}(s)\\
{G_n}(s, \boldsymbol \beta)\\
{M_n}(\boldsymbol \beta, \boldsymbol u)\\
\end{array}\right)=:
    \frac{1}{n}\sum_{i}^n \boldsymbol \psi\left(y^*_{i}, x_{i}, t_{i},\boldsymbol \theta\right)=0,
\end{align}
where $\boldsymbol \theta^\mathsf{T} = (s, \boldsymbol \beta^\mathsf{T}, \boldsymbol u^\mathsf{T})$, $\boldsymbol u^\mathsf{T} = (u_{11}, u_{10}, u_{01}, u_{11})$
and
$$
{M_n}(\boldsymbol u, \boldsymbol \beta) 
:=\frac{1}{n}\sum_{i=1}^n\left(\begin{array}{c} 
 y^*_i(u_{11}-g_1(x_i, \boldsymbol \beta))\\
 (1-y^*_i)(u_{10}-g_1(x_i,\boldsymbol \beta))\\
 y^*_i(u_{01}-g_0(x_i, \boldsymbol \beta))\\
 (1-y^*_i)(u_{00}-g_0(x_i, \boldsymbol \beta))\\
\end{array}\right).
$$
Denoting the resulting estimator by $\hat {\boldsymbol\theta}=\left(\hat s, \,\hat{\boldsymbol \beta}^\mathsf{T}, \hat{\boldsymbol u}^\mathsf{T}\right)^\mathsf{T}$, we have the following theorem.

\begin{Theorem}
 \label{th: theoryglm} Let $\boldsymbol{\theta}_0$ denote the true parameter values for $\boldsymbol{\theta}$. If the true index function $\eta$ has the linear form $\eta(t,\boldsymbol{x}) = a_0+ a_1 t+\boldsymbol{b}^\mathsf{T}\boldsymbol x$, then under some regularity conditions given in Appendix A of Supplementary Materials, we have that $\boldsymbol{\hat{\theta}}$ is consistent for $\boldsymbol \theta_0$, and
$$
\sqrt{n}(\hat{\boldsymbol \theta}-\boldsymbol{\theta}
_0) \overset{d}\to{N}(0,\mathbf{V}({\boldsymbol{\theta}_0})),
$$ 
where the limiting variance of $\sqrt{n}\boldsymbol{\hat \theta}$ can be written as the sandwich matrix form: $\mathbf{V}({\boldsymbol{\theta}_0})=\mathbf{H}({\boldsymbol{\theta}_0})^{-1} \mathbf{B}\big\{\mathbf{H}({\boldsymbol{\theta}_0})^{-1}\big\}^\mathsf{T}$ with $\mathbf{B}({\boldsymbol{\theta}_0})=\mathbb{E}\left\{\boldsymbol \psi(Y^*, X, T, \boldsymbol{\theta}_0) \boldsymbol \psi(Y^*, X, T, \boldsymbol{\theta}_0)^\mathsf{T}\right|S=1\}$ and $\mathbf{H}({\boldsymbol{\theta}_0})=\mathbb{E}\{\partial \boldsymbol \psi(Y^*, X, T, \boldsymbol{\theta}_0) / \partial \boldsymbol{\theta}_0^\mathsf{T}|S=1\}$.
\end{Theorem}

Utilizing Theorem \ref{th: theoryglm}, we can estimate $\tau$ by $\hat{\tau}_{\operatorname{GLM}} = \boldsymbol{q}^\mathsf{T} \hat{\boldsymbol{\theta}}$, where $\boldsymbol{q}^\mathsf{T} = (\boldsymbol{0}^\mathsf{T}, \boldsymbol{c}^\mathsf{T})$, $\boldsymbol{c}^\mathsf{T}=(v^*, v^*-1, v^*, v^*-1)$. Subsequently, applying the Slutsky Theorem, we obtain the asymptotic normality of $\hat{\tau}_{\operatorname{GLM}}$:
$$
\sqrt{n}(\hat{\tau}_{\operatorname{GLM}}-\tau) \overset{d}\to{N}(0,\boldsymbol q^\mathsf{T}\mathbf{V}( {\boldsymbol \theta}_0)\boldsymbol q).
$$ 
The covariance matrix $\mathbf{V}({\boldsymbol \theta}_0)$ can be estimated by $\hat{\mathbf{V}}({\boldsymbol \theta}_0) = \hat{\mathbf{H}}^{-1} \hat{\mathbf{B}}\big\{\hat{\mathbf{H}}^{-1}\big\}^\mathsf{T}$, where \newline
$\hat{\mathbf{H}} =-\frac{1}{n}\sum_{i=1}^n{\boldsymbol \psi}'(y_i^*, x_i, t_i, \hat{\boldsymbol \theta})$ and
$\hat{\mathbf{B}}=\frac{1}{n}\sum_{i=1}^n \boldsymbol \psi(y_i^*, x_i, t_i, \hat{\boldsymbol \theta}) \boldsymbol \psi(y_i^*, x_i, t_i, \hat{\boldsymbol \theta})^\mathsf{T}$.
This allows us to make statistical inference regarding $\tau$.

\subsection{A generalized additive model based estimator}

In real-world studies, the linearity assumption of $\eta(t, \boldsymbol{x})$ is often violated. To enhance the robustness of our method, we employ the generalized additive model \citep{a51, a32} to capture the nonlinear characteristics of $\eta(t, \boldsymbol{x})$. This approach enables us to develop an improved estimator that is resilient to model misspecification.

To begin with, we denote by $\tilde \eta(t_i,\boldsymbol x_i)$ the true index of the $i$-th individual and $\eta(t_i,\boldsymbol x_i)$ a working index. We assume $\tilde \eta(t_i,\boldsymbol x_i)$ has the following additive form:
\begin{align}
    \tilde \eta(t_i,\boldsymbol x_{i}) =  a t_i + \sum_{j=1}^D \tilde \eta_j(x_{ij}),\ j = 1, ..., D,
    \label{eq:tilde eta}
\end{align}
where $x_{ij}$ is the $j$-th covariate of $i$-th sample. We also assume that $\mathbb E[\tilde \eta_j(X_{ij})] = 0$ for $j = 1, \dots, D$ to ensure the identifiability of $\tilde \eta_j$. 
We approximate $\tilde \eta_j(x_{ij})$ by the  following B-spline model:
$$
\eta_j(x_{ij})=\sum_{k=-p+1}^{K_n} B_k^{p}(x_{ij}) b_{k, j}, \  j=1, \ldots, D,
$$
where $B_k^{p}(x)$ is the $p$-th B-spline functions defined recursively $\left(k=-p+1, \ldots, K_n\right)$ \citep{b4}.
Here $K_n$ is the number of knots, $p$ is the degree of B-spline, and $b_{k, j}$'s are unknown parameters. 
To simplify notation, we denote $B_k^{p}(x)$ as $B_k(x)$, unless we explicitly state the degree of the B-spline. 
Our primary focus henceforth will be on the $p$-th B-spline. We model the index of the $i$-th sample as 
$$\eta(t_i, \boldsymbol x_i) =a t_i+\sum_{j=1}^{D}\sum_{k=-p+1}^{K_n} B_k(x_{ij}) b_{k, j}= a t_i+\sum_{j=1}^D \boldsymbol{B}(x_{ij})^\mathsf{T}\boldsymbol b_j = \boldsymbol Z_i \boldsymbol \beta,$$
where $\boldsymbol Z_i = (\boldsymbol B(x_{i1})^\mathsf{T},\ldots,\boldsymbol B(x_{iD})^\mathsf{T},t_i)$, $\boldsymbol{B}(x)=\left(B_{-p+1}(x),\ldots,B_{K_n}(x)\right)^{\mathsf{T}}$, $\boldsymbol \beta^{\mathsf{T}} = (\boldsymbol b^{\mathsf{T}}, a)$, $\boldsymbol b^{\mathsf{T}} = (\boldsymbol b_1^{\mathsf{T}},\ldots, \boldsymbol b_D^{\mathsf{T}})$,
and $\boldsymbol{b}_j = \{b_{k,j}\}_{k=-p+1}^{K_n}, j = 1,\dots, D$. Note that the dimension of $\boldsymbol{\beta}$ is $D(K_n+p)+1$.

The observed log-likelihood function based on the above B-spline approximation shares the same form as (\ref{eq:likelihood}), except that $\eta$ is no longer the true index function $\tilde{\eta}$. We then utilize  the ridge-corrected penalized log-likelihood function proposed by \cite{a32} and \cite{a31}:
\begin{align}
\label{eq:plll}
\ell_{\operatorname{rp},n}\left(\boldsymbol{\beta}, \lambda_{n}, \gamma_n\right) =\ell_n\left(\boldsymbol{\beta}\right)-\sum_{j=1}^D \frac{\lambda_{jn}}{2 n} \boldsymbol{b}_j^{\mathsf{T}} \Delta_m^{\mathsf{T}} \Delta_m \boldsymbol{b}_j
-\frac{\gamma_n}{2 n}\sum_{j=1}^D\boldsymbol{b}_j^{\mathsf{T}} \boldsymbol{b}_j,
\end{align}
where $\lambda_n = \{\lambda_{jn}\}_{j=1}^D$ and $\gamma_n$ are tuning parameters.
$\Delta_m$ is the $m$-th order difference matrix \citep{b5}. The spline parameters are subject to penalization, where the first penalty term is a usual trick in the penalized spline estimators to prevent the estimate from wriggling when the spline dimension $D(K_n + p)$ is large, and the second penalty term aims to ensure the nonsingularity of the Hessian matrix of $\ell_{\operatorname{rp},n}\left(\boldsymbol{\beta}, \lambda_n, \gamma_n\right)$. 
 The score functions are obtained with
\begin{align}
{G_{\operatorname{rp},n}}\left(\boldsymbol{\beta}, \lambda_n, \gamma_n\right)=\frac{\partial \ell_{\operatorname{rp},n}\left(\boldsymbol{\beta}, \lambda_n, \gamma_n\right)}{\partial \boldsymbol{\beta}}\label{eq: gn}.
\end{align}

Similar to the operations in Section \ref{sub:GLM}, we can obtain an estimator of $\tau$ by applying Steps 0-4 with ${G_n}$ in Step 2 being replaced by ${G_{\operatorname{rp}, n}}$. Also, we write Steps 1-3 in a form of estimating equations:
\begin{align}
  \left(
\begin{array}{c}\ {S_n}(s)\\
{G_{\operatorname{rp},n}}(\boldsymbol{\beta}, s, \lambda_n, \gamma_n)\\
{M_n}(\boldsymbol u, \boldsymbol \beta)\end{array}
\right)
=:\frac{1}{n}\sum_{i=1}^n \boldsymbol \psi\left(y^*_{i}, x_{i}, t_{i},\boldsymbol \theta, \lambda_n, \gamma_n \right)
=0\label{eq:psi2}. 
\end{align}
Denote the resulting estimators as $\hat {\boldsymbol\theta}=\left(\hat s, \,\hat{\boldsymbol \beta}^\mathsf{T}, \hat{\boldsymbol u}^\mathsf{T}\right)^\mathsf{T}$. Note that, unlike (\ref{eq:psi1}), the dimension of $\boldsymbol \psi$ in (\ref{eq:psi2}) is not fixed, which increases with the sample size $n$. Similarly, The ATE $\tau$ can be estimated by $\hat \tau_{\operatorname{GAM}} = \boldsymbol q ^\mathsf{T}\hat {\boldsymbol\theta}$.

Before stating asymptotic properties of $\hat \tau_{\operatorname{GAM}}$, we introduce some notations.
Let $\boldsymbol{\theta_0}=(s_0, \boldsymbol{\beta_0^{\mathsf{T}}},\boldsymbol u_0^{\mathsf{T}})^{\mathsf{T}}$, 
where $s_0$ is the solution of $\mathbb E\left[S_n(s)|S=1\right]=0$, 
$\boldsymbol{\beta}_0=\underset{\boldsymbol{\beta}}{\operatorname{argmin}}\mathbb E\left[\log \frac{f\left(Y,  \boldsymbol{X}, \tilde \eta\right)}{f\left(Y , \boldsymbol{X}, \boldsymbol{\beta}\right)}|S=1\right]$ is the best spline approximation of the true index function $\tilde \eta(t,\boldsymbol x)$ based on Kullback-Leibler measure and $\boldsymbol u_0$ is the solution of $\mathbb E[{M_n}(\boldsymbol u, \boldsymbol\beta_0)|S=1]=0$.
 Let $\tau$ denote the true value of ATE. We have the following asymptotic results for $\tau_{\operatorname{GAM}}$.

\begin{Theorem} If the true index function $\tilde \eta$ obeys the additive form as (\ref{eq:tilde eta}), then under some regularity conditions in Appendix A of Supplementary Materials, we have that $\hat \tau_{\operatorname{GAM}}$ is consistent for $\tau$, and $\hat \tau_{\operatorname{GAM}} -\tau$ is asymptotically normal with asymptotic mean $\boldsymbol{\operatorname{Bias}}(\hat \tau_{\operatorname{GAM}})$ (refer to (A.14) in Appendix A for an explicit expression), 
and asymptotic covariance $\mathbf{V}(\hat \tau_{\operatorname{GAM}})=\frac{1}{n} \boldsymbol q^\mathsf{T}\tilde{\mathbf{H}}(\lambda_n)^{-1}\tilde {\mathbf{B}}\big\{ \tilde {\mathbf{H}}(\lambda_n)^{-1}\big\}^\mathsf{T}\boldsymbol q$, where
$$\tilde {\mathbf{H}}(\lambda_n)= \mathbb{E}\left\{\tilde{\boldsymbol{\psi}'}(Y^*, X, T, \boldsymbol \theta_0,\lambda_n, \gamma_n=0) |S=1 \right\},$$
$$\tilde {\mathbf{B}}=\mathbb{E}\left\{\tilde{ \boldsymbol{\psi}}(Y^*, X, T,  \boldsymbol\theta_0,\lambda_n=0, \gamma_n=0) \tilde{ \boldsymbol{\psi}}(Y^*, X, T,  \boldsymbol\theta_0,\lambda_n=0, \gamma_n=0)^\mathsf{T} |S=1 \right\}.$$
Refer to (A.5) and (A.6) in Appendix A for explicit expressions of $\tilde \psi$ and $\tilde \psi'$, respectively.
Furthermore, $\boldsymbol{\operatorname{Bias}}(\hat \tau_{\operatorname{GAM}})=O(n^{-(p+1)/(2p+3)})$ and $\mathbf{V}(\hat \tau_{\operatorname{GAM}})=O(n^{-2(p+1)/(2p+3)})$.
\label{th: theorygam}
\end{Theorem}

\textbf{Remark 1:} Theorem \ref{th: theorygam} demonstrates that $\hat{\tau}_{\operatorname{GAM}}$ is $n^{-(p+1)/(2p+3)}$-consistent and asymptotic normal, and the asymptotic order of $\hat{\tau}_{\operatorname{GAM}}$'s mean squared error is $O(n^{-2(p+1)/(2p+3)})$. These results coincide with those of \cite{a31}. 

\textbf{Remark 2:} If the true index follows a linear form, then $\hat \tau_{\operatorname{GLM}}$ proposed in Section \ref{sub:GLM} is $n^{-1/2}$-consistent. However, $\hat \tau_{\operatorname{GLM}}$ is generally sensitive to the linear assumption. On the other hand, $\hat \tau_{\operatorname{GAM}}$ has a much wider applicability, subject to a lower efficiency in terms of convergence rate.


\textbf{Remark 3:} 
In large sample cases, $\mathbf{V}(\hat \tau_{\operatorname{GAM}})$ can be consistently approximated by \newline $\hat{\mathbf{V}}(\hat \tau_{\operatorname{GAM}})=\frac{1}{n}\boldsymbol{q}^{\mathsf T}\hat{\tilde {\mathbf{H}}}^{-1} \hat{\tilde {\mathbf{B}}}\big\{\hat{\tilde {\mathbf{H}}}^{-1}\big\}^\mathsf{T}\boldsymbol{q}$, with $\hat{\tilde {\mathbf{H}}} =-\frac{1}{n}\sum_{i=1}^n{\boldsymbol \psi}'(y_i^*, x_i, t_i, \hat{\boldsymbol \theta}, \lambda_n, \gamma_n=0)$ and\newline $\hat{\tilde {\mathbf{B}}}=\frac{1}{n}\sum_{i=1}^n\boldsymbol \psi(y_i^*, x_i, t_i, \hat{\boldsymbol \theta},\lambda_n=0, \gamma_n=0) \boldsymbol \psi(y_i^*, x_i, t_i, \hat{\boldsymbol \theta},\lambda_n=0, \gamma_n=0)^\mathsf{T}$. Statistical inference of $\tau$ can be made according to the asymptotic normality of $\hat \tau_{\operatorname{GAM}}$.

\section{Simulaltion studies}
\label{s:simulation}
In this section, we evaluate the finite-sample performance of our proposed  GLM-EE and GAM-EE methods through simulations. 
\subsection{Data generation}
The data generation process consists of two steps: data pool creation and case-control sample selection. We start by creating a data pool representing the target population, where each patient's true disease status and diagnosis status are generated. We independently sample two continuous covariates, $X_1$ and $X_2$, from a standard normal and uniform distribution, respectively. Another discrete covariate, $U$, is sampled from a Bernoulli distribution with $\mathbb P(U=1)=0.5$. The treatment indicator $T$ is sampled from a Bernoulli distribution with $\mathbb P(T=1|X_1, X_2, U) = \text{expit}(1+0.1X_1-0.1X_2-0.5U)$.
To demonstrate the utility of our methods in both linear and nonlinear settings, we consider the following outcome models:
$$
\begin{aligned}
&\textbf{M1:}\quad \tilde \eta = a_0 - 2T-U-0.5X_1+X_2;\\
&\textbf{M2:}\quad \tilde \eta= a_0 - 2T-U- \sin(3\pi X_1) +(3(X_2-0.5))^3;\\
&\textbf{M3:}\quad \tilde \eta= a_0 - 2T-U- \exp(2X_1)-\sin(3\pi X_2)X_2;\\
&\textbf{M4:}\quad \tilde \eta= a_0 - 2T-U- \exp(2X_1) + (3(X_2-0.5))^3+X_1X_2.
\end{aligned}
$$
\textbf{M1} is a typical linear model. \textbf{M2-M3} are non-linear but still follows the additive form in (\ref{eq:tilde eta}). \textbf{M4} violates both linear and additive assumptions.
In all the models, the intercept term $a_0$ is set based on a predetermined disease prevalence $v$. 
The true outcome $Y$ is sampled from the Bernoulli distribution with success probability ${\operatorname{expit}(\tilde\eta)}$. 
We then generate the observed outcome $Y^*$ based on $Y$ with conditional probability $p_{10}$ and $p_{01}$. Thus we construct a data pool to simulate a population of size 1000,000.
We then randomly sample $n/2$ cases and $n/2$ controls from the data pool based on the observed outcome $Y^*$'s, but with only $Y^*, T, U, X_1, X_2$ kept in the subsequent analysis.

We fix $p_{01}=0$ as it is usually very small in practice. We consider various combinations of $v$, $p_{10}$, and $n$. That is,  $v = 0.001,\ 0.01, \text{ and } 0.1$, 
$p_{01}=0,\ 0.2, \text{ and } 0.4$, $n = 500 \text{ and } 2000$. 
 For each combination of $v$ and outcome model, the true $\tau$ are calculated through Monte Carlo integration.
 When applying the GAM-EE method, the number of knots for $X_1$ and $X_2$ is set to be $K_n = 10$, and the value of $\lambda_n$ is selected based on the Bayesian information criterion as in \cite{a32}, and $\gamma_n$ is set to be 0.1.
 For each combination of $v$, $p_{10}$, $p_{01}$, $n$, and outcome model, we repeat simulations for 500 times. 

\subsection{Debias capacity of GLM-EE and GAM-EE}

We first evaluate the debiasing capability of our proposed methods in model \textbf{M1}, where the true index has a linear form. Along with the standard GLM-EE and GAM-EE methods, we also consider three naive estimators based on the GLM-EE method and three naive estimators based on the GAM-EE method.
These naive estimators are obtained by applying the GLM-EE and GAM-EE methods but intentionally ignoring the information of measurement and selection (i.e., manually fix $p_{10} = 0$ and $s=1$ when applying the GLM-EE and GAM-EE methods). 
The EE methods ignoring measurement information, selection information, and both information are denoted as ``naive 3", ``naive 2" and ``naive 1", respectively. 
We also consider the IPTW method as a comparison.

\begin{figure}
\centering
\includegraphics[width = 16cm, height = 5cm]{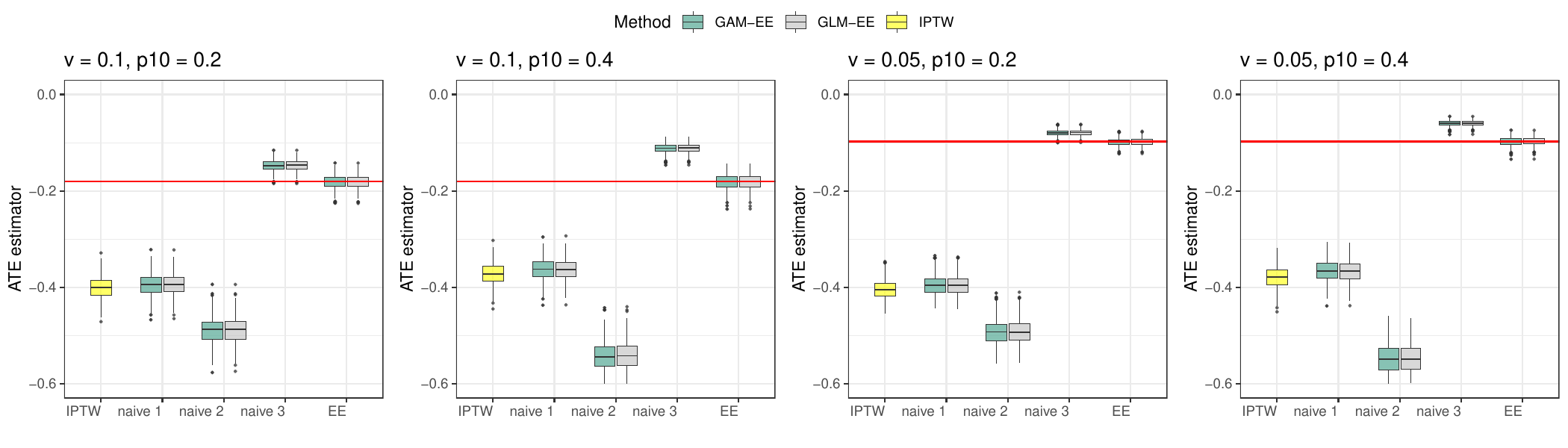}
\caption{Estimators of $\tau$ under model \textbf{M1}. The red line is the true value of $\tau$. naive 3 estimators are the EE methods ignoring the measurement information; naive 2 estimators are the EE methods ignoring the selection information; naive 1 estimators are the EE methods ignoring both selection and measurement information.}
\label{f:M1}
\end{figure}

Figure \ref{f:M1} depicts the box plots of the ATE estimators. The standard GLM-EE and GAM-EE estimators' box covers the true $\tau$ (represented by the red line), and the empirical means closely align with the true $\tau$.
Conversely, the naive and IPTW estimators exhibit obvious bias, as their boxes fail to cover the true $\tau$. The biases are particularly large for IPTW, naive 1, and naive 2, and they are much bigger than naive 3. This observation suggests that the biases are primarily due to sampling bias instead of measurement error.

To further compare the performance of standard GAM-EE and GLM-EE methods, we present the relative biases, root mean squared errors, coverage probabilities of $\hat \tau_{\operatorname{GLM}}$ and $\hat \tau_{\operatorname{GAM}}$ in Table \ref{t:table 1}. Both $\hat \tau_{\operatorname{GLM}}$ and $\hat \tau_{\operatorname{GAM}}$ produce fairly small empirical biases and reasonable coverage probabilities close to the nominal level of 95\%. Furthermore, $\hat \tau_{\operatorname{GAM}}$ has slightly bigger RMSEs than $\hat \tau_{\operatorname{GLM}}$ since GAM-EE requires to estimate more parameters than GLM-EE.

\begin{table}
\caption{Summary of simulation results for \textbf{M1}.}
\label{t:table 1}
\begin{center}
\resizebox{\textwidth}{!}{
\begin{tabular}{llllllllllllll}
\hline
\midrule
       &         & \multicolumn{6}{c}{$n$=500}                           & \multicolumn{6}{c}{$n$=2000}                          \\ \cmidrule(lr){3-8} \cmidrule(lr){9-14}  
       &         & \multicolumn{3}{c}{GAM-EE} & \multicolumn{3}{c}{GLM-EE} & \multicolumn{3}{c}{GAM-EE} & \multicolumn{3}{c}{GLM-EE} \\ \cmidrule(lr){3-5} \cmidrule(lr){6-8} \cmidrule(lr){9-11} \cmidrule(lr){12-14} 
       $p_{10}$&     $v$  & Rbias   & RMSE   & CP    & Rbias   & RMSE   & CP    & Rbias   & RMSE   & CP    & Rbias   & RMSE   & CP    \\ \midrule
       $0$ & 0.001 & 1.72     & 0.023   & 95.6  & 1.51     & 0.022   & 95.0  & -0.14    & 0.011   & 94.4  & -0.19    & 0.011   & 94.8  \\
       & 0.01  & 1.60     & 0.218   & 95.4  & 1.49     & 0.214   & 95.2  & 0.58     & 0.105   & 95.2  & 0.48     & 0.104   & 95.2  \\
       & 0.1   & 1.02     & 1.602   & 94.2  & 0.84     & 1.598   & 94.2  & 0.04    & 0.781   & 96.0  & -0.02    & 0.780   & 96.2  \\
       $0.2$  & 0.001 & 1.58     & 0.022   & 95.4  & 1.50     & 0.022   & 95.8  & 0.40     & 0.010   & 94.8  & 0.37     & 0.010   & 94.6  \\
       & 0.01  & 1.61     & 0.215   & 95.6  & 1.40     & 0.212   & 96.2  & 0.82     & 0.103   & 95.8  & 0.74     & 0.102   & 95.4  \\
       & 0.1   & 1.44     & 1.695   & 95.0  & 1.31     & 1.683   & 95.2  & 0.19     & 0.819   & 95.2  & 0.18     & 0.817   & 95.6  \\
        $0.4$  & 0.001 & 1.93     & 0.022   & 95.8  & 1.70     & 0.022   & 96.6  & 0.64     & 0.012   & 93.4  & 0.60     & 0.012   & 92.8  \\
       & 0.01  & 2.19     & 0.237   & 94.2  & 1.95     & 0.231   & 94.6  & 0.86     & 0.108   & 93.6  & 0.80     & 0.107   & 94.0  \\
       & 0.1   & 0.57     & 1.711   & 96.0  & 0.57     & 1.712   & 95.6  & 0.28     & 0.819   & 95.8  & 0.26     & 0.818   & 96.2  \\ \midrule\hline
\end{tabular}
}
\begin{tablenotes}
        \footnotesize
        \item[1] Rbias, relative bias (\%); RMSE, root mean squared error ($\times$ 1000); CP, coverage probability (\%).
\end{tablenotes}
\end{center}
\end{table}

\subsection{Robustness of GLM-EE and GAM-EE}
To evaluate the robustness of our methods, we conduct a detailed comparison between the standard GLM-EE method ($\hat \tau_{\operatorname{GLM}}$) and the standard GAM-EE method ($\hat \tau_{\operatorname{GAM}}$) in different nonlinear model settings.  
Tables \ref{t:table 2} summarize the results of model \textbf{M2} and \textbf{M3}, where the GLM-EE method suffers from the problem of model misspecification. 
In the simulation situations, $\hat \tau_{\operatorname{GLM}}$ produces systematic biases, depending on the prevalence $v$ (the lower the prevalence, the larger the bias).
 On the other hand, $\hat \tau_{\operatorname{GAM}}$ produces consistently smaller empirical biases and RMSEs than $\hat \tau_{\operatorname{GLM}}$, especially in model \textbf{M3}. 
 The coverage probabilities of $\hat \tau_{\operatorname{GAM}}$ are also close to the nominal level of 95\%.
 This demonstrates the high performance of the GAM-EE method in nonlinear but additive settings. 
Table \ref{t:table 4} shows the results for model \textbf{M4}, where both GLM-EE and GAM-EE methods suffer from model misspecification problems, but $\hat \tau_{\operatorname{GAM}}$ has smaller biases and RMSEs in general.

\begin{table}
\caption{Summary of simulation results for \textbf{M2} and \textbf{M3}.}
\label{t:table 2}
\begin{center}
\resizebox{\textwidth}{!}{
\begin{tabular}{lllllllllllllll}
\hline
\midrule
       &&         & \multicolumn{6}{c}{$n$=500}                         & \multicolumn{6}{c}{$n$=2000}                        \\ \cmidrule(lr){4-9} \cmidrule(lr){10-15}  
       &&         & \multicolumn{3}{c}{GAM-EE} & \multicolumn{3}{c}{GLM-EE} & \multicolumn{3}{c}{GAM-EE} & \multicolumn{3}{c}{GLM-EE} \\ \cmidrule(lr){4-6} \cmidrule(lr){7-9} \cmidrule(lr){10-12} \cmidrule(lr){13-15}
      model &$p_{10}$ &        $v$ & Rbias   & RMSE   & CP   & Rbias   & RMSE   & CP   & Rbias   & RMSE   & CP   & Rbias   & RMSE   & CP   \\  \midrule
       \textbf{M2}&$0$ & 0.001 & -0.22  & 0.027  & 95.6 & -6.04 & 0.027  & 88.6 & -1.92   & 0.016  & 91.0   & 5.59   & 0.020   & 92.6 \\
       && 0.01  & 0.44   & 0.236  & 95.4 & -1.84  & 0.241  & 93.8 & -1.16  & 0.134  & 94.6 & 5.02   & 0.167  & 90.0   \\
       && 0.1   & 0.85   & 1.559  & 96.0   & 4.76   & 1.816  & 92.4 & -0.13  & 0.749  & 94.8 &-0.05  & 0.764  & 94.6 \\ 
       &$0.2$  & 0.001 & 0.19    & 0.027  & 95.8 & -5.06  & 0.026  & 90.0   & -1.10  & 0.017  & 94.0   & 5.79   & 0.021  & 91.0   \\
       && 0.01  & 0.73   & 0.249  & 94.0   & -1.52  & 0.242  & 94.2 & -0.64  & 0.130   & 95.2 & 6.11   & 0.179  & 86.6 \\
       && 0.1   & 0.94   & 1.658  & 94.2 & 4.93   & 1.916  & 92.8 & -0.49  & 0.755  & 95.0   & 0.72   & 0.748  & 96.6 \\
       &$0.4$  & 0.001 & 0.18   & 0.028  & 94.8 & -5.26  & 0.027  & 88.6 & -1.25   & 0.016  & 94.2 & 6.13   & 0.020   & 92.4 \\
       && 0.01  & 1.36   & 0.257  & 93.6 & -1.07  & 0.250   & 94.0   & -1.00  & 0.127  & 94.4 & 5.99   & 0.173  & 90.4 \\
       && 0.1   & 1.00   & 1.521  & 96.8 & 5.40   & 1.830   & 95.2 & -0.70  & 0.802  & 94.0   & 1.00   & 0.829  & 93.4 \\\midrule
\textbf{M3}&$0$ & 0.001 & 0.52     & 0.025   & 95.4  & 17.94    & 0.048   & 83.4  & 0.10     & 0.012   & 94.2  & 16.30    & 0.037   & 34.2  \\
       && 0.01  & 0.33     & 0.241   & 95.8  & 12.27    & 0.358   & 86.0  & -0.16    & 0.118   & 94.2  & 10.71    & 0.251   & 58.4  \\
       && 0.1   & -0.00     & 1.587   & 94.0  & -1.75     & 1.529   & 93.6  & -0.08    & 0.804   & 94.4  & -2.77    & 0.871   & 90.6  \\ 
       &$0.2$  & 0.001 & 0.41     & 0.025   & 95.4  & 18.05    & 0.049   & 86.8  & -0.09    & 0.012   & 94.8  & 16.14    & 0.036   & 37.0  \\
       && 0.01  & 0.19     & 0.231   & 94.8  & 12.78    & 0.357   & 88.6  & -0.67    & 0.115   & 94.6  & 10.90    & 0.253   & 55.2  \\
       && 0.1   & -0.46    & 1.609   & 96.4  & -1.01    & 1.534   & 95.6  & -0.29    & 0.836   & 94.6  & -1.67    & 0.838   & 92.0  \\
       &$0.4$  & 0.001 & 0.22     & 0.025   & 93.4  & 17.65    & 0.048   & 86.0  & 0.03     & 0.012   & 95.4  & 16.35    & 0.037   & 34.4  \\
       && 0.01  & 0.13     & 0.228   & 95.2  & 12.93    & 0.356   & 87.6  & 0.06     & 0.112   & 95.6  & 12.00    & 0.270   & 52.4  \\
       && 0.1   & -1.71    & 1.853   & 93.0  & -1.22    & 1.740   & 93.2  & -0.15    & 0.855   & 94.8  & -0.83    & 0.786   & 95.2  \\\midrule\hline
\end{tabular}
}
\begin{tablenotes}
        \footnotesize
        \item[1] Rbias, relative bias (\%); RMSE, root mean squared error ($\times$ 1000); CP, coverage probability (\%).
\end{tablenotes}
\end{center}
\end{table}

\begin{table}
\caption{Summary of simulation results for \textbf{M4}.}
\label{t:table 4}
\begin{center}
\resizebox{\textwidth}{!}{

\begin{tabular}{llllllllllllll}
\hline
\midrule
       &         & \multicolumn{6}{c}{$n$=500}                           & \multicolumn{6}{c}{$n$=2000}                          \\ \cmidrule(lr){3-8} \cmidrule(lr){9-14} 
       &         & \multicolumn{3}{c}{GAM-EE} & \multicolumn{3}{c}{GLM-EE} & \multicolumn{3}{c}{GAM-EE} & \multicolumn{3}{c}{GLM-EE} \\ \cmidrule(lr){3-5} \cmidrule(lr){6-8} \cmidrule(lr){9-11} \cmidrule(lr){12-14} 
       $p_{10}$&      $v$   & Rbias   & RMSE   & CP    & Rbias   & RMSE   & CP    & Rbias   & RMSE   & CP    & Rbias   & RMSE   & CP    \\ \midrule
       $0$ & 0.001 & -0.68    & 0.031   & 91.8  & 6.27     & 0.033   & 95.2  & 0.16     & 0.014   & 96.0  & 5.69     & 0.018   & 90.8  \\
       & 0.01  & -1.54    & 0.248   & 93.8  & 5.96     & 0.278  & 95.0  & -0.20    & 0.121   & 95.6  & 5.46     & 0.164   & 88.6  \\
       & 0.1   & -1.09     & 1.549   & 94.0  & 1.49     & 1.550   & 94.0  & -1.06    & 0.757   & 93.8  & -0.76    & 0.752  & 95.4 \\
       $0.2$  & 0.001 & 0.13     & 0.029   & 94.4  & 6.88     & 0.032   & 97.4  & 0.54     & 0.014   & 95.4  & 5.74     & 0.019   & 91.8  \\
       & 0.01  & -1.02    & 0.254   & 93.6  & 7.16     & 0.304   & 93.8  & -0.01    & 0.123   & 95.0  & 5.71     & 0.168   & 88.2  \\
       & 0.1   & -1.35    & 1.535   & 94.0  & 2.68     & 1.665   & 93.8  & -0.83    & 0.784   & 94.6  & 2.32     & 0.851   & 94.4  \\ 
        $0.4$  & 0.001 & 0.20     & 0.030   & 94.2  & 7.40     & 0.034   & 96.2  & 0.02     & 0.014   & 95.2  & 5.38     & 0.018   & 92.0  \\
       & 0.01  & -0.89    & 0.254   & 94.2  & 7.47     & 0.309   & 93.8  & -0.27    & 0.127  & 95.0  & 6.00     & 0.173   & 87.0  \\
       & 0.1   & -2.10    & 1.553   & 95.0  & 3.04     & 1.654   & 95.8  & -1.30    & 0.779   & 95.2  & 2.60     & 0.863   & 95.6  \\
       \midrule
\hline
\end{tabular}
}
\begin{tablenotes}
        \footnotesize
        \item[1] Rbias, relative bias (\%); RMSE, root mean squared error ($\times$ 1000); CP, coverage probability (\%).
      \end{tablenotes}
\end{center}
\end{table}

Overall, The GAM-EE method outperforms the GLM-EE method in nonlinear settings and loses little statistical efficiency compared with the GLM-EE method in linear settings. The above results support the theoretical results established in Section \ref{s:model}. 
For scenarios where $p_{01}>0$, as discussed in Section \ref{s:model}, the estimate of $\tau$ is not stable if $v$ is rather small. We increase the sample sizes to $n = 3000$ and only consider four combinations of $p_{01}$ and $v$ (i.e, $p_{01} = 0.03, \,0.06$, and $v = 0.05,\, 0.1$). Tables \ref{t:table s1} in Appendix B of Supplementary Materials summarize the corresponding results,  showing the same behaviors as scenarios with $p_{01} = 0$.

\section{Real data analysis}
\label{s:real data}

In this section, we apply the GAM-EE and GLM-EE methods to a real-world example. We aim to analyze the effect of alcohol intake on the risk of developing gout. We use data from the UK BioBank database, a large-scale prospective cohort study including 502,543 volunteer participants aged 37 to 73 years from UK between 2007 and 2010. We collected information on the treatment (alcohol intake), the observed outcome (gout diagnosis status), and covariates including education level, ethnicity, diet score (summarized score of diet habits), BMI, physical exercise, TDI (Townsend deprivation index), age, and household income. After eliminating the missing data and limiting our sample to only males, we obtained a target population of 136,741 subjects (refer to Table \ref{t:table s2} in Appendix B of Supplementary Materials for detailed information).
Within this population, 3.85\% subjects are diagnosed with gout ($v^*=3.85\%$), but the true disease prevalence $v$ is unknown. However, if we know the values of false positive rate $p_{01}$ and false negative rate $p_{10}$, we can calculate the true disease prevalence by $v = ({v^*-p_{01}})/({1-p_{10}-p_{01}})$ according to (\ref{eq: vstar}). 
We apply our proposed GLM-EE and GAM-EE methods to the dataset. When applying the GAM-EE method, the number of knots $K_n$ is fixed to be 5. The value of $\lambda_n$ is selected based on the Bayesian information criterion from a candidate sequence ranging from 1 to 20, and $\gamma_n$ is set to be 0.1. 

First, we aim to extend the discussion in Section \ref{s:simulation} with the purpose of evaluating the validity of our proposed EE methods, leveraging the full dataset. The corresponding results can be regarded as benchmarks. It is important to mention that while the full dataset does not have the bias sampling problem, it still suffers from measurement error problems. We draw a case-control subsample from the full dataset based on the diagnosed status, with the same case and control size of 2500. The subsample suffers from both selection and mismeasurement biases. We then apply GAM-EE and GLM-EE methods to the subsample. This process is repeated for 500 times. Figure \ref{f:figure5.2} in Appendix B of Supplementary Materials depicts the box plots of our estimators.
The results given by GAM-EE method are fairly close to the corresponding benchmarks, with differences ranging from $1\times10^{-7}$ to $1\times 10^{-2}$. On the other hand, the results given by GLM-EE method deviate from the corresponding benchmark, with differences ranging from $1\times10^{-6}$ to $2\times 10^{-2}$. The standard errors of the two methods are close to each other. These results indicate that GAM-EE method
is more robust than GLM-EE method in this example.

\begin{figure}
\centering
\includegraphics[width = 17cm,height = 6cm]{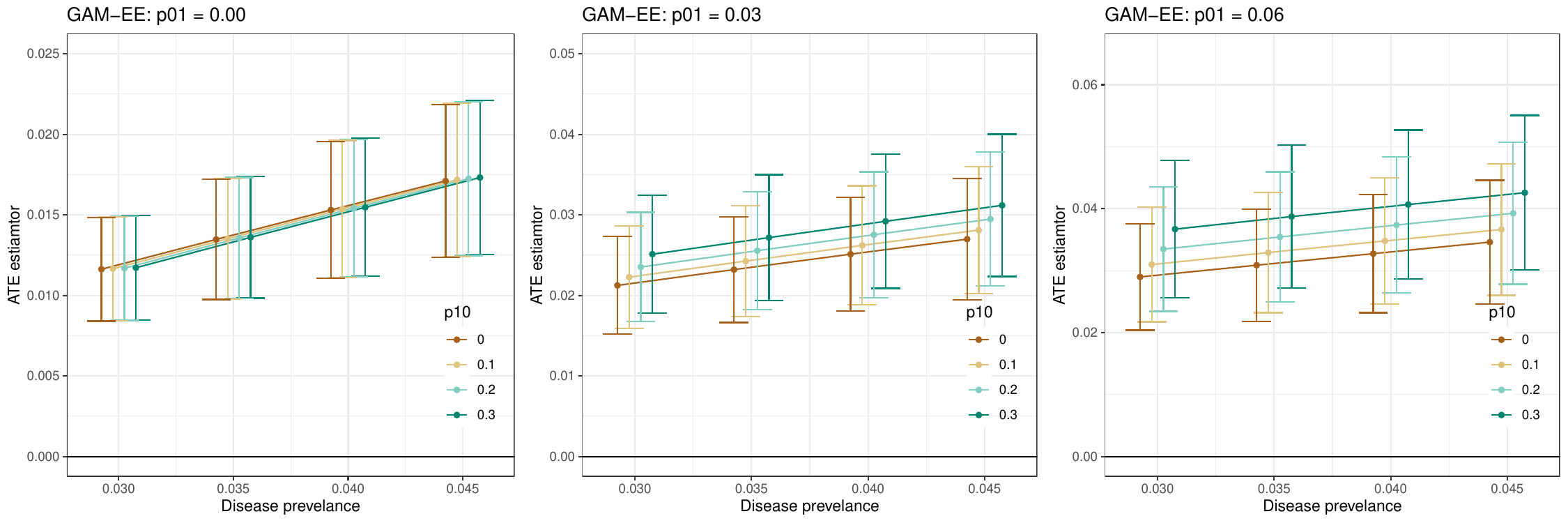}
\caption{$\hat \tau_{\operatorname{GAM}}$ and its 95\% confidence interval in the sensitivity analysis of the UK biobank dataset.}
\label{f:figure6.3}
\end{figure}

Second, we demonstrate the practical utility of our methods in real-world research by conducting a sensitivity analysis. This time we only have a case-control subsample and the real disease status is unobserved. Based on literature review \citep{a34, a33} and experts experience, we determine possible ranges of disease prevalence $v \in (0.030, 0.045)$, false positive rate $p_{10}\in(10\%, 30\%)$, and false negative rate $p_{01} \in (0\%, 6\%)$. 
We select several breakpoints within these ranges, apply our methods, and summarize the results in Figure \ref{f:figure6.3}. Evidently, within the possible ranges of $v$, $p_{10}$, and $p_{01}$, the estimated $\tau$ is significantly greater than 0, in terms of 95\% confidence intervals. The median ATE ranges from 0.01 to 0.04, depending on the specification of $v$, $p_{01}$ and $p_{10}$.
Therefore, we conclude that alcohol intake has a significant positive ATE on the risk of developing gout.

\section{Discussion}

\label{s:discuss}

This paper presents novel methods for addressing an analytical challenge that arises when conducting causal inference in the context of outcome-dependent sampling (ODS) design with measurement error. In such scenarios, ordinary ATE estimators are susceptible to selection and mismeasurement biases. 
Our proposed GLM-EE and GAM-EE methods leverage additional information from the target population on disease prevalence and mismeasurement rates to address the biases.
The effectiveness of our methods for eliminating the influence of ODS and mismeasurement appears to be robust to the specification of outcome models in simulation studies. We also provide practical guidance for conducting sensitivity analysis in real-world ODS studies. 
We apply our methods to the UK Biobank dataset to estimate ATE of alcohol intake on gout. Our methods demonstrated promising performance in this application. 

Our methods focus on estimating ATE, although they can readily be extended to estimate other causal effect measures of interest, such as the causal risk ratio and the causal odds ratio. Furthermore, although we consider a scenario with two treatment options, our methods can be generalized to multiple treatment arms.

As discussed in Section \ref{s:model}, the adjusted link function may be quite flat in cases where $v$ is small and $p_{01}>0$. This can lead to instability in solving the estimating equations, necessitating a large sample size to ensure convergence. However, The increase in sample size will highly increase the computation time, especially when the dimension of covariates or the size of the knots is big. As a result, when $p_{01}>0$, we only consider scenarios where the disease prevalence $v$ is not small. This limitation may restrict the generalizability of our methods and we leave this interesting topic as a future work to explore.

Overall, our proposed methods provide a valuable tool for addressing the analytical challenges associated with causal inference in the presence of ODS and measurement error. Our methods offer a practical and effective means in obtaining unbiased estimates of ATE, even when the outcome model is not linear.


\backmatter

\section*{Acknowledgements}

The work of HZ is partly supported by the National Natural Science Foundation of China (7209121, 12171451). This research has been conducted using the UK Biobank resource (application number 96744), subject to a data transfer agreement.

\section*{Dada availability statement}

The data that support the findings in this paper can be obtained from the  UK  Biobank  (\url{http://www.ukbiobank.ac.uk}).  

\section*{Supplementary Materials}

Appendix A (referenced in Section~\ref{s:identification}-\ref{s:model}) and Appendix B (referenced in Sections~\ref{s:model}-\ref{s:real data}) are available.\vspace*{-8pt}
\bibliographystyle{biom} 
\bibliography{biomsample.bib}

\begin{thebibliography}{}

\bibitem[\protect\citeauthoryear{Angrist, Imbens, and Rubin}{Angrist
  et~al.}{1996}]{a3}
Angrist, J.~D., Imbens, G.~W., and Rubin, D.~B. (1996).
\newblock Identification of causal effects using instrumental variables.
\newblock {\em Journal of the American Statistical Association} {\bf 91,}
  444--455.

\bibitem[\protect\citeauthoryear{Angrist and Krueger}{Angrist and
  Krueger}{2001}]{a4}
Angrist, J.~D. and Krueger, A.~B. (2001).
\newblock Instrumental variables and the search for identification: From supply
  and demand to natural experiments.
\newblock {\em Journal of Economic Perspectives} {\bf 15,} 69--85.

\bibitem[\protect\citeauthoryear{Bareinboim and Pearl}{Bareinboim and
  Pearl}{2016}]{a6}
Bareinboim, E. and Pearl, J. (2016).
\newblock Causal inference and the data-fusion problem.
\newblock {\em Proceedings of the National Academy of Sciences} {\bf 113,}
  7345--7352.

\bibitem[\protect\citeauthoryear{Barrow and Smith}{Barrow and
  Smith}{1978}]{a38}
Barrow, D. and Smith, P. (1978).
\newblock Asymptotic properties of best l\textsubscript{2}[0, 1] approximation
  by splines with variable knots.
\newblock {\em Quarterly of Applied Mathematics} {\bf 36,} 293--304.

\bibitem[\protect\citeauthoryear{Beesley and Mukherjee}{Beesley and
  Mukherjee}{2022}]{a28}
Beesley, L.~J. and Mukherjee, B. (2022).
\newblock Statistical inference for association studies using electronic health
  records: handling both selection bias and outcome misclassification.
\newblock {\em Biometrics} {\bf 78,} 214--226.

\bibitem[\protect\citeauthoryear{Breslow and Holubkov}{Breslow and
  Holubkov}{1997}]{a8}
Breslow, N.~E. and Holubkov, R. (1997).
\newblock Maximum likelihood estimation of logistic regression parameters under
  two-phase, outcome-dependent sampling.
\newblock {\em Journal of the Royal Statistical Society: Series B (Statistical
  Methodology)} {\bf 59,} 447--461.

\bibitem[\protect\citeauthoryear{Copeland, Checkoway, McMichael, and
  Holbrook}{Copeland et~al.}{1977}]{a18}
Copeland, K.~T., Checkoway, H., McMichael, A.~J., and Holbrook, R.~H. (1977).
\newblock Bias due to misclassification in the estimation of relative risk.
\newblock {\em American Journal of Epidemiology} {\bf 105,} 488--495.

\bibitem[\protect\citeauthoryear{Czado and Santner}{Czado and
  Santner}{1992}]{a52}
Czado, C. and Santner, T.~J. (1992).
\newblock The effect of link misspecification on binary regression inference.
\newblock {\em Journal of Statistical Planning and Inference} {\bf 33,}
  213--231.

\bibitem[\protect\citeauthoryear{De~Boor}{De~Boor}{1978}]{b4}
De~Boor, C. (1978).
\newblock {\em A Practical Guide to Splines}, volume~27.
\newblock Springer, New York.

\bibitem[\protect\citeauthoryear{Dierckx}{Dierckx}{1995}]{b5}
Dierckx, P. (1995).
\newblock {\em Curve and Surface Fitting With Splines}.
\newblock Oxford University Press, Oxford.

\bibitem[\protect\citeauthoryear{Fox, Lash, and Greenland}{Fox
  et~al.}{2005}]{a15}
Fox, M.~P., Lash, T.~L., and Greenland, S. (2005).
\newblock A method to automate probabilistic sensitivity analyses of
  misclassified binary variables.
\newblock {\em International Journal of Epidemiology} {\bf 34,} 1370--1376.

\bibitem[\protect\citeauthoryear{Fox, MacLehose, and Lash}{Fox
  et~al.}{2022}]{b3}
Fox, M.~P., MacLehose, R.~F., and Lash, T.~L. (2022).
\newblock {\em Applying Quantitative Bias Analysis to Epidemiologic Data}.
\newblock Springer, New York.

\bibitem[\protect\citeauthoryear{Gabriel, Sachs, and Sj{\"o}lander}{Gabriel
  et~al.}{2022}]{a5}
Gabriel, E.~E., Sachs, M.~C., and Sj{\"o}lander, A. (2022).
\newblock Causal bounds for outcome-dependent sampling in observational
  studies.
\newblock {\em Journal of the American Statistical Association} {\bf 117,}
  939--950.

\bibitem[\protect\citeauthoryear{Hastie and Tibshirani}{Hastie and
  Tibshirani}{1987}]{a51}
Hastie, T. and Tibshirani, R. (1987).
\newblock Generalized additive models: Some applications.
\newblock {\em Journal of the American Statistical Association} {\bf 82,}
  371--386.

\bibitem[\protect\citeauthoryear{Jurek, Maldonado, and Greenland}{Jurek
  et~al.}{2013}]{a24}
Jurek, A.~M., Maldonado, G., and Greenland, S. (2013).
\newblock Adjusting for outcome misclassification: the importance of accounting
  for case-control sampling and other forms of outcome-related selection.
\newblock {\em Annals of Epidemiology} {\bf 23,} 129--135.

\bibitem[\protect\citeauthoryear{Kiefer, Diekhoff, Hermann, Stroux, Mews,
  Blobel, Hamm, and Hermann}{Kiefer et~al.}{2016}]{a33}
Kiefer, T., Diekhoff, T., Hermann, S., Stroux, A., Mews, J., Blobel, J., Hamm,
  B., and Hermann, K.-G.~A. (2016).
\newblock Single source dual-energy computed tomography in the diagnosis of
  gout: Diagnostic reliability in comparison to digital radiography and
  conventional computed tomography of the feet.
\newblock {\em European Journal of Radiology} {\bf 85,} 1829--1834.

\bibitem[\protect\citeauthoryear{L.~Penning~de Vries and
  Groenwold}{L.~Penning~de Vries and Groenwold}{2022}]{a12}
L.~Penning~de Vries, B.~B. and Groenwold, R. H.~H. (2022).
\newblock Identification of causal effects in case-control studies.
\newblock {\em BMC Medical Research Methodology} {\bf 22,} 7.

\bibitem[\protect\citeauthoryear{Lash and Fink}{Lash and Fink}{2003}]{a14}
Lash, T.~L. and Fink, A.~K. (2003).
\newblock Semi-automated sensitivity analysis to assess systematic errors in
  observational data.
\newblock {\em Epidemiology} {\bf 14,} 451--458.

\bibitem[\protect\citeauthoryear{Lyles and Lin}{Lyles and Lin}{2010}]{a16}
Lyles, R.~H. and Lin, J. (2010).
\newblock Sensitivity analysis for misclassification in logistic regression via
  likelihood methods and predictive value weighting.
\newblock {\em Statistics in Medicine} {\bf 29,} 2297--2309.

\bibitem[\protect\citeauthoryear{M{\aa}nsson, Joffe, Sun, and
  Hennessy}{M{\aa}nsson et~al.}{2007}]{a10}
M{\aa}nsson, R., Joffe, M.~M., Sun, W., and Hennessy, S. (2007).
\newblock On the estimation and use of propensity scores in case-control and
  case-cohort studies.
\newblock {\em American Journal of Epidemiology} {\bf 166,} 332--339.

\bibitem[\protect\citeauthoryear{Marx and Eilers}{Marx and Eilers}{1998}]{a32}
Marx, B.~D. and Eilers, P.~H. (1998).
\newblock Direct generalized additive modeling with penalized likelihood.
\newblock {\em Computational Statistics and Data Analysis} {\bf 28,} 193--209.

\bibitem[\protect\citeauthoryear{Neuhaus}{Neuhaus}{1999}]{a19}
Neuhaus, J.~M. (1999).
\newblock Bias and efficiency loss due to misclassified responses in binary
  regression.
\newblock {\em Biometrika} {\bf 86,} 843--855.

\bibitem[\protect\citeauthoryear{Robins}{Robins}{1999}]{a9}
Robins, J.~M. (1999).
\newblock Choice as an alternative to control in observational studies:
  comment.
\newblock {\em Statistical Science} {\bf 14,} 281--293.

\bibitem[\protect\citeauthoryear{Rosenbaum and Rubin}{Rosenbaum and
  Rubin}{1983}]{a1}
Rosenbaum, P.~R. and Rubin, D.~B. (1983).
\newblock {The central role of the propensity score in observational studies
  for causal effects}.
\newblock {\em Biometrika} {\bf 70,} 41--55.

\bibitem[\protect\citeauthoryear{Rubin}{Rubin}{2005}]{a30}
Rubin, D.~B. (2005).
\newblock Causal inference using potential outcomes.
\newblock {\em Journal of the American Statistical Association} {\bf 100,}
  322--331.

\bibitem[\protect\citeauthoryear{Rubin and Thomas}{Rubin and Thomas}{2000}]{a2}
Rubin, D.~B. and Thomas, N. (2000).
\newblock Combining propensity score matching with additional adjustments for
  prognostic covariates.
\newblock {\em Journal of the American Statistical Association} {\bf 95,}
  573--585.

\bibitem[\protect\citeauthoryear{Schlesselman}{Schlesselman}{1982}]{b1}
Schlesselman, J.~J. (1982).
\newblock {\em Case-control Studies: Design, Conduct, Analysis}, volume~2.
\newblock Oxford University Press, Oxford.

\bibitem[\protect\citeauthoryear{Shu and Yi}{Shu and Yi}{2019a}]{a20}
Shu, D. and Yi, G.~Y. (2019a).
\newblock Causal inference with measurement error in outcomes: Bias analysis
  and estimation methods.
\newblock {\em Statistical Methods in Medical Research} {\bf 28,} 2049--2068.

\bibitem[\protect\citeauthoryear{Shu and Yi}{Shu and Yi}{2019b}]{a22}
Shu, D. and Yi, G.~Y. (2019b).
\newblock Weighted causal inference methods with mismeasured covariates and
  misclassified outcomes.
\newblock {\em Statistics in Medicine} {\bf 38,} 1835--1854.

\bibitem[\protect\citeauthoryear{Shu and Yi}{Shu and Yi}{2020}]{a21}
Shu, D. and Yi, G.~Y. (2020).
\newblock Causal inference with noisy data: Bias analysis and estimation
  approaches to simultaneously addressing missingness and misclassification in
  binary outcomes.
\newblock {\em Statistics in Medicine} {\bf 39,} 456--468.

\bibitem[\protect\citeauthoryear{Van~der Laan}{Van~der Laan}{2008}]{a13}
Van~der Laan, M.~J. (2008).
\newblock Estimation based on case-control designs with known prevalence
  probability.
\newblock {\em The International Journal of Biostatistics} {\bf 4,} 1--40.

\bibitem[\protect\citeauthoryear{Van~der Laan and Rose}{Van~der Laan and
  Rose}{2011}]{b2}
Van~der Laan, M.~J. and Rose, S. (2011).
\newblock {\em Targeted Learning: Causal Inference for Observational and
  Experimental Data}, volume~4.
\newblock Springer, New York.

\bibitem[\protect\citeauthoryear{V{\'a}zquez-Mellado, Hern{\'a}ndez-Cuevas,
  Alvarez-Hern{\'a}ndez, Ventura-Rios, Pel{\'a}ez-Ballestas, Casasola-Vargas,
  Garc{\'\i}a-M{\'e}ndez, and Burgos-Vargas}{V{\'a}zquez-Mellado
  et~al.}{2012}]{a34}
V{\'a}zquez-Mellado, J., Hern{\'a}ndez-Cuevas, C.~B., Alvarez-Hern{\'a}ndez,
  E., Ventura-Rios, L., Pel{\'a}ez-Ballestas, I., Casasola-Vargas, J.,
  Garc{\'\i}a-M{\'e}ndez, S., and Burgos-Vargas, R. (2012).
\newblock The diagnostic value of the proposal for clinical gout diagnosis
  (\uppercase\expandafter{CGD}).
\newblock {\em Clinical Rheumatology} {\bf 31,} 429--434.

\bibitem[\protect\citeauthoryear{Wacholder, Silverman, McLaughlin, and
  Mandel}{Wacholder et~al.}{1992}]{a7}
Wacholder, S., Silverman, D.~T., McLaughlin, J.~K., and Mandel, J.~S. (1992).
\newblock Selection of controls in case-control studies:.
  \uppercase\expandafter{\romannumeral2}. types of controls.
\newblock {\em American Journal of Epidemiology} {\bf 135,} 1029--1041.

\bibitem[\protect\citeauthoryear{Yoshida and Naito}{Yoshida and
  Naito}{2014}]{a31}
Yoshida, T. and Naito, K. (2014).
\newblock Asymptotics for penalised splines in generalised additive models.
\newblock {\em Journal of Nonparametric Statistics} {\bf 26,} 269--289.

\end{thebibliography}
\appendix

\section{A. Technical Details of Sections 2-3}


\subsection{A.1 Proof of Lemma 2.1}

Utilizing the law of total probability, we have
$$
\begin{aligned}
    \mathbb P(Y^*=1) = &\mathbb P(Y^*=1|Y=1)\mathbb P(Y=1) + \mathbb P(Y^*=1|Y=0)\mathbb P(Y=0)\\
    =&\mathbb P(Y^*=1|Y=1)\mathbb P(Y=1) + \mathbb P(Y^*=1|Y=0)(1-\mathbb P(Y=1))\\
    =&(\mathbb P(Y^*=1|Y=1)-\mathbb P(Y^*=1|Y=0))\mathbb P(Y=1) + \mathbb P(Y^*=1|Y=0).
\end{aligned}
$$ 
That is $v^*=(p_{11}-p_{01})v+p_{01}.$

Let $s_{01} = \mathbb P(S=1|Y^*=1)$, $s_{11} = \mathbb P(S=1|Y^*=0)$.
Applying Bayes rule and Assumption 4, we have
$$
\begin{aligned}
s_{11} =\mathbb P(S=1|Y^*=1)=\mathbb P(Y^*=1|S=1)*\mathbb P(S=1)/\mathbb P(Y^*=1),\\
s_{01} =\mathbb P(S=1|Y^*=0)=\mathbb P(Y^*=0|S=1)*\mathbb P(S=1)/\mathbb P(Y^*=0),
\end{aligned}
$$
thus $s = s_{11}/s_{01} = \frac{\mathbb P(Y^*=1|S=1)/v^*}{\mathbb P(Y^*=0|S=1)/(1-v^*)}$.
Applying Bayes rule, 
\begin{align}
&\mathbb P(Y^*=1|X,T=1,S=1)\nonumber \\=&\frac{\mathbb P(Y^*=1,S=1|X,T=1)}{\mathbb P(S=1|X,T=1)}\nonumber \\
=&\frac{\mathbb P(S=1|X,Y^*=1)\mathbb P(Y^*=1|X,T=1)}{\mathbb P(S=1|X,Y^*=1)\mathbb P(Y^*=1|X,T=1)+\mathbb P(S=1|X,Y^*=0)\mathbb P(Y^*=0|X,T=1)}\nonumber \\
=&\frac{\mathbb P(S=1|Y^*=1)\mathbb P(Y^*=1|X,T=1)}{\mathbb P(S=1|Y^*=1)\mathbb P(Y^*=1|X,T=1)+\mathbb P(S=1|Y^*=0)\mathbb P(Y^*=0|X,T=1)}\nonumber.
\end{align}
With Assumption 5, it follows:
\begin{align}
\label{eq: Ap2}
\mathbb P(Y^*=1|X,T=1)=\frac{s_{01}\mathbb P(Y^*=1|X,T=1,S=1)}{s_{11}-\mathbb P(Y^*=1|X,T=1,S=1)(s_{11}-s_{01})}.
\end{align}
We also have
$$
\begin{aligned}
\mathbb P(Y^*=1|X,T=1)=&\mathbb P(Y^*=1|X,Y=1,T=1)\mathbb P(Y=1|X,T=1)\\&+\mathbb P(Y^*=1|X,Y=0,T=1)\mathbb P(Y=0|X,T=1)
\\=&\mathbb P(Y^*=1|Y=1)\mathbb P(Y=1|X,T=1)\\&+\mathbb P(Y^*=1|Y=0)\mathbb P(Y=0|X,T=1),
\end{aligned}
$$
which follows that
\begin{align}
\label{eq: Ap3}
\mathbb P(Y=1|X,T=1)=\frac{\mathbb P(Y^*=1|X,T=1)-p_{01}}{p_{11}-p_{01}}.
\end{align}
With (\ref{eq: Ap2}) and (\ref{eq: Ap3}), we obtain
$$
\begin{aligned}
g_1(X)=\mathbb{E}[Y|X,T=1]=\frac{1}{p_{11}-p_{01}}\left(\frac{s_{01}g^*_1(X)}{s_{11}-g^*_1(X)(s_{11}-s_{01})}-p_{01}\right)
\end{aligned}
$$
Analogously, we have
$$
g_0(X)=\mathbb{E}[Y|X,T=1]=\frac{1}{p_{11}-p_{01}}\left(\frac{s_{01}g^*_0(X)}{s_{11}-g^*_0(X)(s_{11}-s_{01})}-p_{01}\right).\
$$

 \newpage

\subsection{A.2 Proof of Theorem 3.1}

\subsubsection{A.2.1 Notations} 
~\\
Let $\boldsymbol{\theta}_0$ denote the true parameter values for $\boldsymbol{\theta}$. 
Let $\boldsymbol{\psi}^{\prime}(\boldsymbol{\theta})$ denote$\partial \boldsymbol \psi(  Y^*, X, T, \boldsymbol{\theta}) / \partial \boldsymbol \theta^{\top}$ 
and $\boldsymbol{\psi}^{\prime \prime}(\boldsymbol{\theta})$ denote $\partial^2 \boldsymbol{\psi}\left(Y^*, X, T,\boldsymbol{\theta} \right) / \partial \boldsymbol{\theta} \partial \boldsymbol{\theta}^{\top}$. 
Let $\mathbf{H}\left(\boldsymbol{\theta}\right)$ denote $\mathbb{E}\left\{\boldsymbol{\psi}'\left(Y^*, X, T,\boldsymbol{\theta}  \right)|S=1\right\}$ and $\mathbf{B}\left(\boldsymbol{\theta}\right)$ denote $\mathbb{E}\left\{\boldsymbol{\psi}\left( Y^*, X, T, \boldsymbol{\theta} \right) \boldsymbol{\psi}\left( Y^*, X, T, \boldsymbol{\theta} \right)^{\top}|S=1\right\}$. 

\subsubsection{A.2.2 Regularity conditions for Theorem 3.1} 
~\\
(A1) Assume that $\boldsymbol{\psi}^{\prime}(\boldsymbol{\theta})$ and $\boldsymbol{\psi}^{\prime \prime}(\boldsymbol{\theta})$ exist and uniformly bounded for all $\boldsymbol{\theta}$ in an open neighborhood of $\boldsymbol{\theta}_0$.~\\
(A2) $\mathbf{H}\left(\boldsymbol{\theta}_0\right)$ and $\mathbf{B}\left(\boldsymbol{\theta}_0\right)$ are positive definite.

\subsubsection{A.2.3 Proof of Theorem 3.1} 
~\\
By standard Z-estimation theory and $\mathbb{E} [\boldsymbol\psi\left(\boldsymbol{\theta}_0\right)|S=1]=0$, a solution $\hat{\boldsymbol{\theta}}$ to the estimating equation $\boldsymbol\psi(\boldsymbol{\theta})=0$, exists, unique, and consistent, i.e., $\hat{\boldsymbol{\theta}} \stackrel{p}{\rightarrow} \boldsymbol{\theta}_0$. To establish asymptotic normality, the estimators $\hat{\boldsymbol{\theta}}$ solve the equation
$$
\mathbf{\Psi}_n(\hat{\boldsymbol{\theta}}) := \frac{1}{n} \sum_{i=1}^n \boldsymbol{\psi}\left( y_i^*, x_i, t_i,\hat{\boldsymbol{\theta}}\right)=\mathbf{0} .
$$

Taking Taylor expansion of $\mathbf{\Psi}_n(\hat{\boldsymbol{\theta}})$, we have
$$
\mathbf{0}=\mathbf{\Psi}_n(\hat{\boldsymbol{\theta}})=\mathbf{\Psi}_n\left(\boldsymbol{\theta}_0\right)+\mathbf{\Psi}^{\prime}_n\left(\boldsymbol{\theta}_0\right)\left(\hat{\boldsymbol{\theta}}-\boldsymbol{\theta}_0\right)+\mathbf{R}_n .
$$
where 
$\mathbf{\Psi}^{\prime}_n\left(\boldsymbol{\theta}_0\right)=-\partial \mathbf{\Psi}_n(\boldsymbol{\theta}) /\left.\partial \boldsymbol{\theta}^{\top}\right|_{\boldsymbol{\theta}=\boldsymbol{\theta}_0}$.
Under regularity conditions specified above,\newline $-\mathbf{\Psi}^{\prime}_n\left(\boldsymbol{\theta}_0\right)\stackrel{p}{\longrightarrow}\mathbf{H}\left(\boldsymbol{\theta}_0\right)$ and $\sqrt{n} \mathbf{R}_n \stackrel{p}{\longrightarrow} 0$ as $n \rightarrow \infty$, so that
$$
\sqrt{n}\left(\hat{\boldsymbol{\theta}}-\boldsymbol{\theta}_0\right) \approx \mathbf{H}^{-1}\left(\boldsymbol{\theta}_0\right) \sqrt{n} \mathbf{\Psi}_n\left(\boldsymbol{\theta}_0\right) .
$$
By applying the Central Limit Theorem,
$$
\sqrt{n}\mathbf{\Psi}_n\left(\boldsymbol{\theta}_0\right) \stackrel{d}{\longrightarrow} N\left\{\mathbf{0}, \mathbf{B}\left(\boldsymbol{\theta}_0\right)\right\} .
$$
Therefore, we have that
$$
\sqrt{n}\left(\hat{\boldsymbol{\theta}}-\boldsymbol{\theta}_0\right) \sim N\left(\mathbf{0}, \mathbf{H}\left(\boldsymbol{\theta}_0\right)^{-1} \mathbf{B}\left(\boldsymbol{\theta}_0\right)\left\{\mathbf{H}\left(\boldsymbol{\theta}_{\mathbf{0}}\right)^{-1}\right\}^{\top}\right) \text { as } n \rightarrow \infty
$$
\newpage
\subsection{A.3 Proof of Theorem 3.2}
\subsubsection{A.3.1 Notations} 
~\\
 As denoted in Section 3, for $i$-th individual, $\boldsymbol Z_i = (\boldsymbol B(x_{i1})^{\top},...,\boldsymbol B(x_{iD})^{\top},T_i)$, where $\boldsymbol{B}(x)=\left(B_{-p+1}(x), \cdots, B_{K_n}(x)\right)^{^{\top}}$, $\boldsymbol \beta^{\top} = (\boldsymbol b^{\top}, a)$, $\boldsymbol b^{\top} = (\boldsymbol b_1^{\top},\dots, \boldsymbol b_D^{\top})$ and $\boldsymbol{b}_j = \{b_{k,j}\}_{k=-p+1}^{K_n}, j = 1,\dots, D$. The explicit expressions for the estimating equations are provided below:
$$
\begin{aligned}
  &\boldsymbol{\Psi}_n\left(\boldsymbol \theta, \lambda_n, \gamma_n \right):=\frac{1}{n}\sum_{i}^n \boldsymbol \psi\left(y_i^*,x_i,t_i,\boldsymbol \theta, \lambda_n, \gamma_n \right) =
\left(
\begin{array}{c}S_n(s)\\
G_{\operatorname{rp},n}(\boldsymbol{\beta}, s, \lambda_n, \gamma_n)\\
M_n( \eta(\boldsymbol \beta),\boldsymbol u)\end{array}
\right),  \\
\text{where }\\
   & S_n(s) = \frac{1}{n} \sum_{i = 1}^n\{s v^*    (1 - y_i^*) - (1 - v^*) y_i^* \} = 0,\\
&G_{\operatorname{rp},n}\left(\boldsymbol{\beta}, s,\lambda_n, \gamma_n\right)=\frac{\partial \ell_{\operatorname{rp},n}\left(\boldsymbol{\beta},s, \lambda_n, \gamma_n\right)}{\partial \boldsymbol{\beta}}=\frac{1}{n}\boldsymbol{Z}^{^{\top}}U(\eta(\boldsymbol{\beta}), s)-\frac{1}{n} Q_m^a\left(\lambda_n\right) \boldsymbol{\beta}-\frac{1}{n}T^a(\gamma_n)  \boldsymbol{\beta}, \\
&M_n( \eta(\boldsymbol \beta),\boldsymbol u) 
=\frac{1}{n}\sum_{i=1}^n\left(\begin{array}{c} 
 y^*_i(u_{11}-\operatorname{expit}(\eta(T = 1, X = \boldsymbol{x_i},\boldsymbol{\beta})))\\
 (1-y^*_i)(u_{10}-\operatorname{expit}(\eta(T = 1, X = \boldsymbol{x_i},\boldsymbol{\beta})))\\
 y^*_i(u_{01}-\operatorname{expit}(\eta(T = 0, X = \boldsymbol{x_i},\boldsymbol{\beta})))\\
 (1-y^*_i)(u_{00}-\operatorname{expit}(\eta(T = 0, X = \boldsymbol{x_i},\boldsymbol{\beta})))\\
\end{array}\right),\\
&U(\eta(\boldsymbol{\beta}),s) = \left\{(Y_i-h(\eta(\boldsymbol{Z}_i,\boldsymbol{\beta})))\frac{h'(\eta(\boldsymbol{Z}_i,\boldsymbol{\beta}))}{h(\eta(\boldsymbol{Z}_i,\boldsymbol{\beta}))(1-h(\eta(\boldsymbol{Z}_i,\boldsymbol{\beta})))}\right\}_{\{i=1,\dots n\}},\\
&Q_m^a(\lambda_n) = 
\left[\begin{array}{cc}Q_m(\lambda_n)&0\\
0&0\end{array}\right],
T^a(\gamma_n) = 
\left[\begin{array}{cc}\gamma_nI&0\\
0&0\end{array}\right], 
Q_m(\lambda_n) = \operatorname{diag}[\lambda_{1n}\Delta_m^{\mathsf{T}} \Delta_m\ldots \lambda_{Dn}\Delta_m^{\mathsf{T}} \Delta_m].
\nonumber
\end{aligned}
$$
Note that only the spline parameters $\boldsymbol b_j, j=1,\dots, D$ are penalized. The Hessian matrix can be written as:

\begin{align}
&\mathbf{H}_n(\boldsymbol{\theta}, \lambda_n, \gamma_n) :=   \boldsymbol{\Psi}_n^{\prime}\left(\boldsymbol \theta, \lambda_n, \gamma_n \right) =
\left[\begin{array}{ccc}\mathbf{H}_{n,1}(s)&0&0\\
\mathbf{H}_{n,2}(\boldsymbol{\beta},s)&\mathbf{H}_{n,3}(\boldsymbol{\beta},s)&0\\
0&\mathbf{H}_{n,4}(\boldsymbol{u},\boldsymbol{\beta})&\mathbf{H}_{n,5}
\end{array}\right],
\end{align} 
\begin{align}
\text{where}\\
&\mathbf{H}_{n,1}\left(s \right)=\frac{\partial S_n\left(s\right)}{\partial s },\quad \mathbf{H}_{n,2}\left(\boldsymbol{\beta},s \right)=\frac{\partial G_{\operatorname{rp},n}\left(\boldsymbol{\beta},s\right)}{\partial s },    \nonumber\\
&\mathbf{H}_{n,3}\left(\boldsymbol{\beta},s \right)=\frac{\partial G_{\operatorname{rp},n}\left(\boldsymbol{\beta},s\right)}{\partial \boldsymbol{\beta} }=-\frac{1}{n} Z^{^{\top}} W(\eta(\boldsymbol{\beta}),s) Z-\frac{1}{n} Q^a_m\left(\lambda_n\right)-\frac{1}{n} T^a(\gamma_n),    \nonumber\\
&\mathbf{H}_{n,4}\left(\boldsymbol{\beta},\boldsymbol{u}\right)=\frac{\partial M_n\left( \eta(\boldsymbol \beta),\boldsymbol u)\right)}{\partial \boldsymbol{\beta} },\quad \mathbf{H}_{n,5}=\frac{\partial M_n\left( \eta(\boldsymbol \beta),\boldsymbol u\right)}{\partial \boldsymbol{u}}.
    \nonumber
\end{align} 
and $W(\eta(\boldsymbol{\beta}),s) = \operatorname {diag}\left[w_i(\eta(\boldsymbol{Z}_i,\boldsymbol{\beta}),s)\right]$ is a $n$-dimensional diagonal weight matrix resulting from the variance function, with
$$
\begin{aligned}
&w_i=\frac{h'(\eta(\boldsymbol{Z}_i,\boldsymbol{\beta}))^2}{V(\eta(\boldsymbol{Z}_i,\boldsymbol{\beta}))} -
\{Y_i - h(\eta(\boldsymbol{Z}_i,\boldsymbol{\beta}))\} \frac{h''(\eta(\boldsymbol{Z}_i,\boldsymbol{\beta}))V(\eta(\boldsymbol{Z}_i,\boldsymbol{\beta})) - h'(\eta(\boldsymbol{Z}_i,\boldsymbol{\beta}))^2V'(\eta(\boldsymbol{Z}_i,\boldsymbol{\beta}))}{V(\eta(\boldsymbol{Z}_i,\boldsymbol{\beta}))^2},\\
&V(\eta(\boldsymbol{Z}_i,\boldsymbol{\beta}))=h(\eta(\boldsymbol{Z}_i,\boldsymbol{\beta}))(1-h(\eta(\boldsymbol{Z}_i,\boldsymbol{\beta}))).
\end{aligned}
$$

 We further define
\begin{align}
  &\tilde{\boldsymbol{\Psi}}_n\left(\boldsymbol \theta, \lambda_n, \gamma_n \right):=\frac{1}{n}\sum_{i}^n \tilde{\boldsymbol \psi}\left(y_i^*,x_i,t_i,\boldsymbol \theta, \lambda_n, \gamma_n \right) :=
\left(
\begin{array}{c}S_n(s)\\
\tilde{G}_{rp,n}(\boldsymbol{\beta}, s, \lambda_n, \gamma_n)\\
\tilde{M}_n(\boldsymbol u)\end{array}
\right),  \\\nonumber
&\tilde{G}_{rp,n}\left(\boldsymbol{\beta}, s,\lambda_n, \gamma_n\right)=\frac{1}{n}\boldsymbol{Z}^{^{\top}}U(\tilde{\eta}, s)-\frac{1}{n} Q_m^a\left(\lambda_n\right) \boldsymbol{\beta}-\frac{1}{n}T^a(\gamma_n)  \boldsymbol{\beta}, \nonumber\\
&\tilde{M}_n( \boldsymbol u) 
=\frac{1}{n}\sum_{i=1}^n\left(\begin{array}{c} 
 y^*_i(u_{11}-\operatorname{expit}(\tilde{\eta}(T = 1, X = \boldsymbol{x_i})))\\
 (1-y^*_i)(u_{10}-\operatorname{expit}(\tilde{\eta}(T = 1, X = \boldsymbol{x_i})))\\
 y^*_i(u_{01}-\operatorname{expit}(\tilde{\eta}(T = 0, X = \boldsymbol{x_i})))\\
 (1-y^*_i)(u_{00}-\operatorname{expit}(\tilde{\eta}(T = 0, X = \boldsymbol{x_i})))\\
\end{array}\right).
\nonumber
\end{align}
The definition of $\tilde
{\mathbf{H}}_n(\boldsymbol{\theta}, \lambda_n, \gamma_n)$ is similar:

\begin{align}
&\tilde{\mathbf{H}}_n(\boldsymbol{\theta}, \lambda_n, \gamma_n) :=   \frac{1}{n}\sum_{i}^n \tilde{\boldsymbol \psi}^{\prime}\left(y_i^*,x_i,t_i,\boldsymbol \theta, \lambda_n, \gamma_n \right)  :=
\left[\begin{array}{ccc}\mathbf{H}_{n,1}(s)&0&0\\
\tilde{\mathbf{H}}_{n,2}(s)&\tilde{\mathbf{H}}_{n,3}(s)&0\\
0&\tilde{\mathbf{H}}_{n,4}&\mathbf{H}_{n,5}
\end{array}\right],
\end{align}
where
\begin{align}
&\tilde{\mathbf{H}}_{n,2}\left(s \right)=\frac{1}{n}\boldsymbol{Z}^{^{\top}} \partial U(\tilde{\eta}, s)/\partial s,  \nonumber\\
&\tilde{\mathbf{H}}_{n,3}\left(s \right)=-\frac{1}{n} \boldsymbol{Z}^{^{\top}}W(\tilde{\eta},s) \boldsymbol{Z}-\frac{1}{n} Q^a_m\left(\lambda_n\right)-\frac{1}{n} T^a(\gamma_n),  \nonumber\\
&\tilde{\mathbf{H}}_{n,4}=\frac{1}{n}\sum_{i=1}^n\left(\begin{array}{c} 
 y^*_i \operatorname{expit}^{\prime}(\tilde{\eta}(T = 1, X = \boldsymbol{x}_i))\boldsymbol{Z}_i^{\top}\\
 (1-y^*_i)\operatorname{expit}^{\prime}(\tilde{\eta}(T = 1, X = \boldsymbol{x}_i))\boldsymbol{Z}_i^{\top}\\
 -y^*_i\operatorname{expit}^{\prime}(\tilde{\eta}(T = 0, X = \boldsymbol{x}_i))\boldsymbol{Z}_i^{\top}\\
 (1-y^*_i)\operatorname{expit}^{\prime}(\tilde{\eta}(T = 0, X = \boldsymbol{x}_i))\boldsymbol{Z}_i^{\top}\\
\end{array}\right).
    \nonumber
\end{align} 
To simplify notation, we rewrite  $\tilde{\mathbf{H}}_n(\boldsymbol{\theta}, \lambda_n, \gamma_n)$ as $\tilde{\mathbf{\Psi}}_n^{\prime}(s, \lambda_n, \gamma_n)$.

Let $\boldsymbol{\theta_0}=(s_0, \boldsymbol{\beta_0^{\top}},\boldsymbol u_0^{\top})^{\top}$, 
where $s_0$ is the solution of $\mathbb E\left[S_n(s)|S=1\right]=0$, \newline
$\boldsymbol{\beta}_0=\left(\boldsymbol{b}_{0}^{^{\top}},a_{0}\right)^{^{\top}}=\underset{\boldsymbol{\beta}}{\operatorname{argmin}}\mathbb E\left[\log \frac{f\left(Y,  \boldsymbol{X}, \tilde \eta\right)}{f\left(Y , \boldsymbol{X}, \boldsymbol{\beta}\right)}|S=1\right]$ is the best spline approximation of true index function $\tilde \eta(t,\boldsymbol x)$ based on Kullback-Leibler measure and 
$\boldsymbol u_0$ is the solution of $\mathbb E[{M_n}(\boldsymbol u, \boldsymbol\beta_0)|S=1]=0$. 
Let $\tau_0 = \mathbb E[\operatorname{expit}(\eta_0(T=1, X))-\operatorname{expit}(\eta_0(T=0, X))]$, where $\eta_0(t,\boldsymbol x) = \boldsymbol Z\boldsymbol \beta_0$.
Define
$$
\tilde {\mathbf{H}}(\lambda_n)= \mathbb{E}\left\{\tilde{\boldsymbol{\Psi}}'_n(\boldsymbol{\theta}_0,\lambda_n, \gamma_n=0) |S=1 \right\},
$$
$$
\tilde {\mathbf{B}}=n\mathbb{E}\left\{\tilde{\boldsymbol{\Psi}}_n(\boldsymbol{\theta}_0, \lambda_n, \gamma_n=0){\boldsymbol{\Psi}}_n(\boldsymbol{\theta}_0, \lambda_n, \gamma_n=0)^{\top} |S=1 \right\}.
$$
Noting that 
$$\tilde {\mathbf{H}}(\lambda_n)=\mathbb{E}\left\{\tilde{\boldsymbol{\psi}'}(Y^*, X, T, \boldsymbol \theta_0,\lambda_n, \gamma_n=0) |S=1 \right\},$$
$$\tilde {\mathbf{B}}=\mathbb{E}\left\{\tilde{ \boldsymbol{\psi}}(Y^*, X, T,  \boldsymbol\theta_0,\lambda_n=0, \gamma_n=0) \tilde{ \boldsymbol{\psi}}(Y^*, X, T,  \boldsymbol\theta_0,\lambda_n=0, \gamma_n=0)^{\top} |S=1 \right\}.$$


\subsubsection{A.3.2 Regularity conditions for Theorem 3.2} 
~\\
To give the asymptotic properties of $\hat \tau_{\operatorname{GAM}}$, we make the following assumptions:

(B1) We assume that the covariate points $\boldsymbol x_i = (x_{i1}, \ldots, x_{iD})$ are distributed according to a density with compact support on $[0,1]^D$, where $[0,1]^D$ is the $D$-variate unit cube.

(B2) For simplicity, the knots for the B-spline basis are equidistantly located so that  $\kappa_k=k / K_n(k=-p+\left.1, \ldots, K_n+p\right)$ with $\kappa_j-\kappa_{j-1}=O(K_n^{-1})$ for $j=-p+1, \ldots, K_n+p$.

(B3) The smoothing parameter is often chosen as $\lambda_{jn} \to \infty$ with $n \to \infty$ because a spline curve often yields overfitting for large $n$. Thus we assume the penalty parameter $\lambda_{jn}(j=1,\ldots,D)$ grow with the sample size with order
$$
\lambda_{jn}=o\left(nK_n^{-1}\right).
$$

(B4) We assume that the dimension of the B-spline basis grows with the sample size with order
$$
K_n=o\left(n^{1 / 2}\right).
$$
and for the non-singularity of $\mathbf{H}_{n,3}(\boldsymbol \theta, \lambda_n, \gamma_n)$, $K_n$ is chosen such that $D(K_n+p)+1<n$.

(B5) The ridge corrected penalty satifies $\gamma = o(\lambda_n K_n^{-m})$, where $\lambda_n = \max_j{\lambda_{jn}}$.


Moreover, the function $\tilde{\eta}_j(j=1,\ldots,D)$ is assumed to be (p+1) times differentiable and expects a finite number of isolated points in [0,1] it is continuously differentiable. This guarantees that the likelihood contributions are all of the same asymptotic order $O_p(1)$.

\subsubsection{A.3.3 Lemmas for Theorem 3.2} 
~\\
Before the proof of theorem 3.2, We give several lemmas as follows:

\textbf{Lemma B1}\quad
Under Assumptions B1-B5, we have
$$
\hat{\boldsymbol \beta} - \boldsymbol{\beta}_0 = -\mathbf{H}_{n,3}(\boldsymbol{\beta}_0, s_0, \lambda_n, \gamma_n)^{-1}G_{\operatorname{rp},n}(\boldsymbol{\beta}_0, s_0, \lambda_n, \gamma_n)+o_p\left(\left\{\left(\frac{\lambda_n K_n^{1-m}}{n}\right)^2+\frac{K_n}{n}\right\} \boldsymbol{1}\right).
$$





\textbf{Lemma B2}\quad
Under Assumptions B1-B5, we have
$$
\tilde\eta\left(\boldsymbol{x}\right)-\eta_0\left(\boldsymbol{x}\right)=b_{ a}\left(\boldsymbol{x}\right)+o\left(K_n^{-(p+1)}\right)
$$
where $b_a(\boldsymbol{x})=\sum_{j=1}^D b_{j, a}\left(x_j\right)$,
\begin{align}
b_{j, a}(x)=\frac{\tilde \eta_j^{(p+1)}(x)}{K_n^{p+1}(p+1) !} \sum_{k=1}^{K_n} I\left(\kappa_{k-1} \leq x<\kappa_k\right) \operatorname{Br}_{p+1}\left(\frac{x-\kappa_{k-1}}{K_n^{-1}}\right),
\label{eq:ba}
\end{align}
and $I(a<x<b)$ is the indicator function of an interval $(a, b)$ and $\operatorname{Br}_p(x)$ is the $p$-th Bernoulli polynomial.

\subsubsection{A.3.4 Proof of Theorem 3.2} 
~\\
To assess the difference between $(\hat{\tau}_{\operatorname{GAM}} - \tau)$, we decompose it into two distinct components: $(\hat{\tau}_{\operatorname{GAM}} - \tau_0)$ and $(\tau_0 - \tau)$ and evaluate them by the following steps.
$\boldsymbol{Step\;1}$, we establish the asymptotic properties of $\hat{\boldsymbol{\theta}} - \boldsymbol{\theta}_0$. 
$\boldsymbol{Step\;2}$, we determine the asymptotic order of $\mathbb{E}(\hat{\boldsymbol{\theta}} - \boldsymbol{\theta}_0|S=1)$ and $\mathbb{V}(\hat{\boldsymbol{\theta}} - \boldsymbol{\theta}_0|S=1)$.
$\boldsymbol{Step\;3}$, we establish the asymptotic properties of $(\hat{\tau}_{\operatorname{GAM}} - \tau_0)$.
$\boldsymbol{Step\;4}$, we evaluate $(\tau - \tau_0)$. 
~\\~\\
$\boldsymbol{Step\;1}$: we fist prove that 
\begin{align}
\hat{\boldsymbol{\theta}}-\boldsymbol{\theta}_0=-\mathbf{H}_n\left(\boldsymbol{\theta}_0, \lambda_n, \gamma_n\right)^{-1} \mathbf{\Psi_n}\left(\boldsymbol{\theta}_0, \lambda_n, \gamma_n\right)+O_p\left(\left\{\left(\frac{\lambda_n K_n^{1-m}}{n}\right)^2+\frac{K_n}{n}\right\} \boldsymbol{1}\right).
\label{eq:A5}
\end{align}

Taking Taylor's expansion of 
$\mathbf{\Psi_n}\left(\hat{\boldsymbol{\theta}}, \lambda_n, \gamma_n\right)$ around
$\boldsymbol{\theta}_0$, we have
$$
\begin{aligned}
0=& \mathbf{\Psi_n}\left(\hat{\boldsymbol{\theta}}, \lambda_n, \gamma_n\right) \\
=& \mathbf{\Psi_n}\left(\boldsymbol{\theta}_0, \lambda_n, \gamma_n\right)+\mathbf{H}_n\left(\boldsymbol{\theta}_0, \lambda_n, \gamma_n\right)\left(\hat{\boldsymbol{\theta}}-\boldsymbol{\theta}_0\right) \\
&+\left\{\mathbf{H}_n\left(\boldsymbol{\theta}_0+\Omega\left(\hat{\boldsymbol{\theta}}-\boldsymbol{\theta}_0\right), \lambda_n, \gamma_n\right)-\mathbf{H}_n\left(\boldsymbol{\theta}_0, \lambda_n, \gamma_n\right)\right\}\left(\hat{\boldsymbol{\theta}}-\boldsymbol{\theta}_0\right),
\end{aligned}
$$ where
$\Omega=\operatorname{diag}\left[\omega_1, \cdots, \omega_{D\left(K_n+p\right)},\omega_{a} \right]$
and $\omega_i \in(0,1)$. 
To simplify notation, the tuning parameters $\lambda_n$ and $\gamma_n$ are omitted in the subsequent proof. By simple calculation, we rewrite the difference:
$$
\begin{aligned}
& \mathbf{H}_n\left(\boldsymbol{\theta}_0+\Omega\left(\hat{\boldsymbol{\theta}}-\boldsymbol{\theta}_0\right)\right)-\mathbf{H}_n\left(\boldsymbol{\theta}_0\right) \\
= & {\left[\begin{array}{ccc} 
0 & 0 & 0 \\
\mathbf{H}_{n,2}\left(\boldsymbol{\theta}_0+\Omega\left(\hat{\boldsymbol{\theta}}-\boldsymbol{\theta}_0\right)\right)-\mathbf{H}_{n,2}\left(\boldsymbol{\theta}_0\right) & \mathbf{H}_{n,3}\left(\boldsymbol{\theta}_0+\Omega\left(\hat{\boldsymbol{\theta}}-\boldsymbol{\theta}_0\right)\right)-\mathbf{H}_{n,3}\left(\boldsymbol{\theta}_0\right) & 0 \\
0 & \mathbf{H}_{n,4}\left(\boldsymbol{\theta}_0+\Omega\left(\hat{\boldsymbol{\theta}}-\boldsymbol{\theta}_0\right)\right)-\mathbf{H}_{n,4}\left(\boldsymbol{\theta}_0\right) & 0
\end{array}\right] } \\
=: & {\left[\begin{array}{ccc}
0 & 0 & 0 \\
\Delta \mathbf{H}_{n,2} & \Delta \mathbf{H}_{n,3} & 0 \\
0 & \Delta \mathbf{H}_{n,4} & 0
\end{array}\right] },
\end{aligned}
$$
 
We also calculate the inverse of $\mathbf{H}_n\left(\boldsymbol{\theta}_0\right)$:
$$
\begin{aligned}
 \mathbf{H}_n\left(\boldsymbol{\theta}_0\right)^{-1} 
=  {\left[\begin{array}{ccc}
\mathbf{H}_{n,1}^{-1} & 0 & 0 \\
-\mathbf{H}_{n,3}^{-1}\mathbf{H}_{n,2}\mathbf{H}_{n,1}^{-1} & \mathbf{H}_{n,3}^{-1} & 0 \\
\mathbf{H}_{n,5}^{-1}\mathbf{H}_{n,4}\mathbf{H}_{n,3}^{-1}\mathbf{H}_{n,2}\mathbf{H}_{n,1}^{-1} & -\mathbf{H}_{n,5}^{-1}\mathbf{H}_{n,4}\mathbf{H}_{n,3}^{-1} & \mathbf{H}_{n,5}^{-1}
\end{array}\right] },
\end{aligned}
$$
Thus we have 
$$
\begin{aligned}
&\hat{\boldsymbol{\theta}}-\boldsymbol{\theta}_0 \\
= & -\mathbf{H}_n\left(\boldsymbol{\theta}_0\right)^{-1} \mathbf{\Psi_n}\left(\boldsymbol{\theta}_0\right)-\mathbf{H}_n\left(\boldsymbol{\theta}_0\right)^{-1}{\left[\begin{array}{ccc}
0 & 0 & 0 \\
\Delta \mathbf{H}_{n,2} & \Delta \mathbf{H}_{n,3} & 0 \\
0 & \Delta \mathbf{H}_{n,4} & 0
\end{array}\right] }(\hat{\boldsymbol{\theta}}-\boldsymbol{\theta}_0) \\
= & -\mathbf{H}_n\left(\boldsymbol{\theta}_0\right)^{-1}\mathbf{\Psi_n}\left(\boldsymbol{\theta}_0\right)
\\& + {\left[\begin{array}{c}
0  \\
-\mathbf{H}_{n,3}^{-1} \Delta \mathbf{H}_{n,2}(\hat{s}-s_0)-\mathbf{H}_{n,3}^{-1} \Delta \mathbf{H}_{n,3}(\boldsymbol{\hat{\beta}}-\boldsymbol{\beta_0}) \\
\mathbf{H}_{n,5}^{-1}\mathbf{H}_{n,4}\mathbf{H}_{n,3}^{-1}\Delta \mathbf{H}_{n,2} (\hat{s}-s_0)-(\mathbf{H}_{n,5}^{-1}\Delta \mathbf{H}_{n,4}-\mathbf{H}_{n,5}^{-1}\mathbf{H}_{n,4}\mathbf{H}_{n,3}^{-1}\Delta \mathbf{H}_{n,3})(\hat{\boldsymbol{\beta}}-\boldsymbol{\beta_0})
\end{array}\right] }.
\end{aligned}
$$
According to Lemma A.1-A.3 given by \cite{a31}, we can easily obtain that
$$
\mathbf{H}_{n,3} = O_p(K_n^{-1}\boldsymbol{1}\boldsymbol{1}^{\top}),
$$
$$
\mathbf{H}_{n,3}^{-1} \Delta \mathbf{H}_{n,3} =o_p\left(\left\{\frac{\lambda_n K_n^{1-m}}{n}+\sqrt{\frac{K_n}{n}}\right\} \boldsymbol{1}\boldsymbol{1}^{\top}\right).
$$
 It is worth noting that $\partial U(\eta(\beta),s)/\partial s$ is continuously differentiable up to the (p+1)-th order with respect to s, and uniformly bounded for $s\ge1$. Through straightforward calculations, it can be demonstrated that
 $$\mathbf{H}_{n,2}=O_p(K_n^{-1} \boldsymbol 1),\ \mathbf{H}_{n,4} = O_p(K_n^{-1}\boldsymbol{1}\boldsymbol{1}^{\top}),$$ 
$$\Delta \mathbf{H}_{n,4} = O_p\left(\left\{\frac{\lambda_n K_n^{1-m}}{n}+\sqrt{\frac{K_n}{n}}\right\}\boldsymbol{1}\boldsymbol{1}^{\top}\right).$$ 
What's more, it can be easily derived that $\hat{s} - s_0 = O_p(1/\sqrt{n})$. Based on all the aforementioned results and Lemma B1, we have
$$
\begin{aligned}
&-\mathbf{H}_{n,3}^{-1} \Delta \mathbf{H}_{n,2}(\hat{s}-s_0)-\mathbf{H}_{n,3}^{-1} \Delta \mathbf{H}_{n,3}(\boldsymbol{\hat{\beta}}-\boldsymbol{\beta_0}) = o_p\left(\left\{\left(\frac{\lambda_n K_n^{1-m}}{n}\right)^2+\frac{K_n}{n}\right\} \boldsymbol{1}\right),
\end{aligned}
$$
$$
\begin{aligned}
&\mathbf{H}_{n,5}^{-1}\mathbf{H}_{n,4}\mathbf{H}_{n,3}^{-1}\Delta \mathbf{H}_{n,2} (\hat{s}-s_0)-(\mathbf{H}_{n,5}^{-1}\Delta \mathbf{H}_{n,4}-\mathbf{H}_{n,5}^{-1}\mathbf{H}_{n,4}\mathbf{H}_{n,3}^{-1}\Delta \mathbf{H}_{n,3})(\hat{\boldsymbol{\beta}}-\boldsymbol{\beta_0})\\&= O_p\left(\left\{\left(\frac{\lambda_n K_n^{1-m}}{n}\right)^2+\frac{K_n}{n}\right\} \boldsymbol{1}\right).
\end{aligned}
$$
The proof of $\boldsymbol{Step\;1}$ is finished.

$\boldsymbol{Step\;2}$: We next calculate the order of $\mathbb{E}(\hat{\boldsymbol{\theta}}-\boldsymbol{\theta}_0|S=1)$ and $\mathbf{V}(\hat{\boldsymbol{\theta}}-\boldsymbol{\theta}_0|S=1)$.

According to LLN, it can be proved that 
$$
\mathbf{H}_n(\boldsymbol{\theta_0})^{-1}=\tilde{\mathbf{H}}(\lambda_n)^{-1}\left(1+o_p\left(\boldsymbol 1 \boldsymbol 1^{\top}\right)\right).
$$
Thus we have
  \begin{align}
\notag
\hat{\boldsymbol{\theta}}-\boldsymbol{\theta}_0& =-\mathbf{H}_n\left(\boldsymbol{\theta}_0\right)^{-1} \mathbf{\Psi}_n\left(\boldsymbol{\theta}_0\right)+O_p\left(\left\{\left(\frac{\lambda_n K_n^{1-m}}{n}\right)^2+\frac{K_n}{n}\right\} \mathbf{1}\right) \\ \notag
& =-\tilde{\mathbf{H}}(\lambda_n)^{-1}\left(1+o_p\left(11^{\top}\right)\right) \mathbf{\Psi}_n\left(\boldsymbol{\theta}_0\right)+O_p\left(\left\{\left(\frac{\lambda_n K_n^{1-m}}{n}\right)^2+\frac{K_n}{n}\right\} \mathbf{1}\right) \\
& =-\tilde{\mathbf{H}}(\lambda_n)^{-1}\mathbf{\Psi}_n\left(\boldsymbol{\theta}_0\right)+o_p\left(\left\{\frac{\lambda_n K_n^{1-m}}{n}+\sqrt{\frac{K_n}{n}}\right\}\mathbf{1}\right). 
\label{eq:A6}\\\notag
\end{align}  
Taking the expectation on both sides of the equation, we obtain
$$
\begin{aligned}
 \mathbb{E}\left(\hat{\boldsymbol{\theta}}-\boldsymbol{\theta}_0 | S = 1\right)&=-\tilde{\mathbf{H}}(\lambda_n)^{-1} \mathbb{E}\left(\mathbf{\Psi}_n\left(\boldsymbol{\theta}_0\right)\right)+o\left(\left\{\frac{\lambda_n K_n^{1-m}}{n}+\sqrt{\frac{K_n}{n}}\right\}\mathbf{1}\right) \\  
&=-\tilde{\mathbf{H}}(\lambda_n)^{-1} \mathbb{E}\left(\left[\begin{array}{c}\mathbf{S}_n(s_0)\\  
\mathbf{G}_{rp,n}(\boldsymbol{\beta_0}, s_0)\\
\mathbf{M}_n(\boldsymbol u_0, \boldsymbol \beta_0)\end{array}
\right]\right)+o\left(\left\{\frac{\lambda_n K_n^{1-m}}{n}+\sqrt{\frac{K_n}{n}}\right\}\mathbf{1}\right)  \\  
& =-\tilde{\mathbf{H}}(\lambda_n)^{-1} \left[\begin{array}{c}
0\\
-\frac{1}{n} Q_m\left(\lambda_n\right) \boldsymbol{\beta}_0-\frac{\gamma_n}{n} \boldsymbol{\beta}_0 \\
0
\end{array}\right]
+o\left(\left\{\frac{\lambda_n K_n^{1-m}}{n}+\sqrt{\frac{K_n}{n}}\right\}\mathbf{1}\right)\\   &=O\left(\left\{\frac{\lambda_n K_n^{1-m}}{n}\right\}\mathbf{1}\right) + o\left(\left\{\sqrt{\frac{K_n}{n}}\right\}\mathbf{1}\right).
\end{aligned}
$$
Here, we have used the fact:
$$
\begin{aligned}
\left(\frac{\gamma_n}{n}\right) \mathbf{\tilde{H}}_{n,3}^{-1} \boldsymbol{\beta}_0 & =o\left(\lambda_n K_n^{1-m} n^{-1} \mathbf{1}\right), \\
\left(\frac{\gamma_n}{n}\right) \mathbf{\tilde{H}}_{n,4}\mathbf{\tilde{H}}_{n,3}^{-1} \boldsymbol{\beta}_0 & =o\left(\lambda_n K_n^{1-m} n^{-1}\right).
\end{aligned}
$$
According to (\ref{eq:A6}) and LLN, we have
$$
\begin{gathered}
\mathbb{V}\left(\mathbf{\Psi}_n(\boldsymbol{\theta_0})|S=1\right)=\frac{1}{n} \tilde {\mathbf{B}}\left(1+o_p\left(\boldsymbol 1 \boldsymbol 1^{\top}\right)\right) ,
\end{gathered}
$$
\begin{align}
\mathbb{V}\left(\hat{\boldsymbol{\theta}}-\boldsymbol{\theta}_0|S=1\right)
&=\frac{1}{n} \tilde{\mathbf{H}}(\lambda_n)^{-1}\tilde {\mathbf{B}}\left\{\tilde{\mathbf{H}}(\lambda_n)^{-1}\right\}^{\top}+o\left(\frac{K_n}{n}\right)=O\left(\frac{K_n}{n}\right),
\end{align}
where
$$
\tilde {\mathbf{B}}=\mathbb{E}\left\{\tilde{ \boldsymbol{\psi}}(Y^*, X, T,\boldsymbol{\theta}_0, \lambda_n=0, \gamma_n=0) \tilde{ \boldsymbol{\psi}}(Y^*, X, T,\boldsymbol{\theta}_0,\lambda_n=0, \gamma_n=0)^{\top}|S=1\right\}.
$$
 ~\\~\\
$\boldsymbol{Step\;3}$: We give the asymptotic normality
of the difference $(\hat{\tau}_{\operatorname{GAM}} - \tau_0)$.

Let
$\boldsymbol{c}=\left(c_{11}, c_{10}, c_{01}, c_{00}\right)^{\top}$ and $\boldsymbol{q} = (\boldsymbol{0}^{\top}, \boldsymbol{c}^{\top})^{\top}$, we have $\hat{\tau}_{\operatorname{GAM}}=\boldsymbol{q}^{\top} \hat{\boldsymbol{\theta}}$. 
Thus applying the CLT and the Slutsky formula we obtain:
$$
\hat{\tau}_{\operatorname{GAM}}-\tau_0 \stackrel{a}{\sim} \mathcal{N}(\operatorname{Bias}_{\lambda}(\hat{\tau}_{\operatorname{GAM}}), \mathbf{V}(\hat{\tau}_{\operatorname{GAM}})),
$$
where 
\begin{align}
\operatorname{Bias}_{\lambda}(\hat{\tau}_{\operatorname{GAM}})=\boldsymbol{q}^{\top} \operatorname{Bias}(\hat{\boldsymbol{\theta}}),
\label{eq:A7}
\end{align}
\begin{align}
\mathbf{V}(\hat{\tau}_{\operatorname{GAM}})=\boldsymbol{q}^{\top} \mathbf{V}(\hat{\boldsymbol{\theta}}) \boldsymbol{q},
\label{eq:A8}
\end{align}
$$
\begin{aligned}
\operatorname{Bias}(\hat{\boldsymbol{\theta}}) = -\tilde{\mathbf{H}}(\lambda_n)^{-1}\left[\begin{array}{c}
0\\
-\frac{1}{n} Q_m\left(\lambda_n\right) \boldsymbol{\beta}_0-\frac{\gamma_n}{n} \boldsymbol{\beta}_0 \\
0
\end{array}\right]=O\left(\frac{\lambda_n K_n^{1-m}}{n}\right)+o\left(\frac{K_n}{n}\right), 
\end{aligned}
$$
$$
\begin{aligned}
\mathbf{V}(\hat{\boldsymbol{\theta}}) = \mathbb{V}\left(\hat{\boldsymbol{\theta}}-\boldsymbol{\theta}_0|S=1\right)
&=\frac{1}{n} \tilde{\mathbf{H}}(\lambda_n)^{-1}\tilde {\mathbf{B}}\left\{\tilde{\mathbf{H}}(\lambda_n)^{-1}\right\}^{\top} = O\left(\frac{K_n}{n}\right).
\end{aligned}
$$
~\\~\\
$\boldsymbol{Step\;4}$: We give the asymptotic normality of the difference $(\tau - \tau_0)$. 

By Taylor expansion and Lemma B2, we have
$$
\begin{aligned}
 \tau-\tau_0 = & \mathbb{E}\left\{\operatorname{expit}\left(\tilde{\eta}(T=1,X)\right)-\operatorname{expit}\left(\tilde{\eta}(T=0,X)\right)\right\} \\
&- \mathbb{E}\left\{\operatorname{expit}\left(\eta_0(T=1,X)\right)-\operatorname{expit}\left( \eta_0(T=0,X)\right)\right\} \\
= & \mathbb{E}\left\{\operatorname{expit}^{\prime}\left(\tilde{\eta}(T=1,X)\right)\left(\tilde{\eta}(T=0,X)-\eta_0(T=0,X)\right)\right\} \\
&-  \mathbb{E}\left\{\operatorname{expit}^{\prime}\left(\tilde{\eta}(T=0,X)\right)\left(\tilde{\eta}(T=0,X)-\eta_0(T=0,X)\right)\right\}\\
&+o\left(\mathbb{E}\left\{\tilde{\eta}(T=0,X)-\eta_0(T=0,X)\right\}\right) \\
= & \mathbb{E}\left\{\left[\operatorname{expit}^{\prime}\left(\tilde{\eta}(T=1,X)\right)-\operatorname{expit}^{\prime}\left(\tilde{\eta}(T=0,X)\right)\right]\left(\tilde{\eta}(T=0,X)-\eta_0(T=0,X)\right)\right\}\\
&+o\left(\mathbb{E}\left\{\tilde{\eta}(T=0,X)-\eta_0(T=0,X)\right\}\right) \\
= & \mathbb{E}\left\{\left[\operatorname{expit}^{\prime}\left(\tilde{\eta}(T=1,X)\right)-\operatorname{expit}^{\prime}\left(\tilde{\eta}(T=0,X)\right)\right] b_a(X)\right\}+o\left(\mathbb{E}\left\{b_a(X)\right\}\right) \\
= & O\left(K_n^{-(p+1)}\right).
\end{aligned}
$$
We denote
\begin{align}
\operatorname{Bias}_a(\hat{\tau}_{\operatorname{GAM}})=\mathbb{E}\left\{\left[\operatorname{expit}^{\prime}\left(\tilde{\eta}(T=1,X)\right)-\operatorname{expit}^{\prime}\left(\tilde{\eta}(T=0, X)\right)\right] b_a(X)\right\}.
\label{eq:A9}
\end{align}
where $b_a(X)$ is given in (\ref{eq:ba}).
Finally we obtain that bias of ($\hat{\tau}_{\operatorname{GAM}}-\tau$), which equals to 
\begin{align}
\operatorname{Bias}(\hat{\tau}_{\operatorname{GAM}}) = \operatorname{Bias}_{a}(\hat{\tau}_{\operatorname{GAM}})+\operatorname{Bias}_{\lambda}(\hat{\tau}_{\operatorname{GAM}}),
\label{eq:A10}
\end{align}
with $\operatorname{Bias}_{\lambda}(\hat{\tau}_{\operatorname{GAM}})$ given in (\ref{eq:A7}) and $\operatorname{Bias}_{a}(\hat{\tau}_{\operatorname{GAM}})$ given in (\ref{eq:A9}).

To conclude, the
bias of $\hat{\tau}_{\operatorname{GAM}}$ can be bounded as 
$$\operatorname{Bias}(\hat{\tau}_{\operatorname{GAM}})=O\left(\lambda_n K_n^{1-m} / n+K_n^{-(p+1)}\right)+o\left({K_n}/{n}\right).$$
Furthermore, 
$$\mathbf{V}(\hat{\tau}_{\operatorname{GAM}})=O\left(K_n/n\right).$$
Thus we have mean squared error of $\hat{\tau}_{\operatorname{GAM}}$:
$$\operatorname{MSE}(\hat{\tau_{\operatorname{GAM}}}) = O\left(K_n/n+K_n^{-2(p+1)}\right).$$
By setting $K_n=O\left(n^{1 /(2 p+3)}\right)$ and $\lambda_n=O\left(n^v\right)$, where $v \leqslant(p+m+1) /(2 p+3)$, we obtain the smallest order of MSE, which is $O\left(n^{-(2 p+2) /(2 p+3)}\right)$.
Thus we finish the proof.

\subsubsection{A.3.5 Proof of Lemmas}
~\\
Next we give the proof of the two lemmas above. 

\textbf{Proof of Lemma B1}: It is worth mentioning that our link function, denoted as $h(\eta)$, exhibits properties similar to the logistic link function. Therefore, the proof for Lemma B1 closely resembles that of Lemma A.4 in the reference paper \cite{a31}, as well as Lemma A.1 to A.3 in \cite{a31}.

\textbf{Proof of Lemma B2}: \cite{a38} showed that there exists $\boldsymbol{\beta}^{*} \in \mathbb{R}^{D\left(K_n+p\right)+1}$ such that
$$
\sup _{\boldsymbol{x} \in(0,1)^D}\left|\tilde{\eta}(\boldsymbol{x})+b_a(\boldsymbol{x})-\boldsymbol{Z}{^{\top}} \boldsymbol{\beta}^{*}\right|=o\left(K_n^{-(p+1)}\right)
$$
Where $\boldsymbol{\beta}^{*}=\left(\left(\boldsymbol{b}^*\right)^{\top}, a^*\right)$.
Let
$$
\eta^*(\boldsymbol{x})=T a^*+\sum_{j=1}^D \eta_j^*\left(x_j\right)=T a^*+\sum_{j=1}^D \boldsymbol{B}\left(x_j\right)^{^{\top}} \boldsymbol{b}_j^*,
$$
and
$$
\eta_0(\boldsymbol{x})=T a_0+\sum_{j=1}^D \eta_{j 0}\left(x_j\right)=T a_0+\sum_{j=1}^D \boldsymbol{B}\left(x_j\right)^{\top} \boldsymbol{b}_{j 0}
$$
Since the asymptotic orders of $\eta_0(\mathbf{x})-\eta^*(\mathbf{x})$ and that of $\boldsymbol{\beta}_0-\boldsymbol{\beta}^{*}$ are the same, if $\boldsymbol{\beta}_0-\boldsymbol{\beta}^{*}=o\left(K_n^{-(p+1)} \mathbf{1}\right)$, we obtain for any $\mathbf{x} \in(0,1)^D,\left|\eta_0(\mathbf{x})-\eta^*(\mathbf{x})\right|=o\left(K_n^{-(p+1)}\right)$ hence Lemma B2 holds.
Now we turn to prove that:
$$
\boldsymbol{\beta}_0-\boldsymbol{\beta}^{*}=o\left(K_n^{-(p+1)} \mathbf{1}\right)
$$
From the definition of $\boldsymbol{\beta}_0$, we have
\begin{align}
\mathbb{E}\left[\log \frac{f\left(Y_1 \mid \boldsymbol{X}, \tilde\eta\right)}{f\left(Y_1 \mid \boldsymbol{X}, \boldsymbol{\beta}_0\right)}|S=1\right] \leq \mathbb{E}\left[\log \frac{f\left(Y_1 \mid \boldsymbol{X}, \tilde\eta\right)}{f\left(Y_1 \mid \boldsymbol{X}, \boldsymbol{\beta}^{*}\right)}|S=1\right]\label{eq: A13}
\end{align}
Since $b_a(x)=O\left(K_n^{-(p+1)}\right)$ and $\frac{h^{\prime}\left(\tilde{\eta}\left(x\right)\right)^2}{V\left(h(\tilde{\eta}(x))\right)}$ is bounded, through the Taylor expansion, we have
\begin{align}
\mathbb{E}\left[\log \frac{f\left(Y_1 \mid \boldsymbol{X}, \tilde{\eta}\right)}{f\left(Y_1 \mid \boldsymbol{X}, \boldsymbol{\beta}^{*}\right)}|S=1\right] & =\mathbb{E}\left\{\mathbb{E}\left[\log \frac{f\left(Y_1 \mid \boldsymbol{X}, \tilde{\eta}\right)}{f\left(Y_1 \mid \boldsymbol{X}, \boldsymbol{\beta}^{*}\right)} \mid \boldsymbol{X},S=1\right]|S=1\right\}\nonumber \\
& =\frac{1}{2} \mathbb{E}\left[\left\{\tilde{\eta}\left(\boldsymbol{X}\right)-\eta^*\left(\boldsymbol{X}\right)\right\}^2 \frac{h^{\prime}\left(\tilde{\eta}\left(\boldsymbol{X}\right)\right)^2}{V\left(h(\tilde{\eta}(\boldsymbol{X}))\right)}\left(1+o_p(1)\right)|S=1\right] \nonumber\\
& =O\left(K_n^{-2(p+1)}\right)\label{eq: A14},
\end{align}
and
\begin{align}
\mathbb{E}\left[\log \frac{f\left(Y_1 \mid \boldsymbol{X}, \tilde{\eta}\right)}{f\left(Y_1 \mid \boldsymbol{X}, \boldsymbol{\beta}_0\right)}|S=1\right] & =\mathbb{E}\left\{\mathbb{E}\left[\log \frac{f\left(Y_1 \mid \boldsymbol{X}, \tilde{\eta}\right)}{f\left(Y_1 \mid \boldsymbol{X}, \boldsymbol{\beta}_0\right)} \mid \boldsymbol{X},S=1\right]|S=1\right\} \nonumber\\
& =\frac{1}{2} \mathbb{E}\left[\left\{\tilde{\eta}\left(\boldsymbol{X}\right)-\eta_0\left(\boldsymbol{X}\right)\right\}^2 \frac{h^{\prime}\left(\tilde{\eta}\left(\boldsymbol{X}\right)\right)^2}{V\left(h(\tilde{\eta}(\boldsymbol{X}))\right)}\left(1+o_p(1)\right)|S=1\right] \nonumber\\
& =\frac{1}{2} \mathbb E\left[\left\{\tilde{\eta}\left(\boldsymbol{X}\right)-\boldsymbol Z^{\top} \boldsymbol{\beta}_0\right\}^2 \boldsymbol{C}|S=1\right]\label{eq: A15},
\end{align}
where $\boldsymbol{C}=\frac{h^{\prime}\left(\tilde{\eta}\left(\boldsymbol X\right)\right)^2}{V\left(h(\tilde{\eta}(\boldsymbol{X}))\right)}\left(1+o_p(1)\right)$. 
With (\ref{eq: A13}-\ref{eq: A15}), we obtain $|\tilde \eta(x)-\eta_0(x)|=o(1)$.

Since $\boldsymbol{\beta}_0$ is the minimiser of $\frac{1}{2} \mathbb{E}\left[\left\{\tilde{\eta}\left(\boldsymbol{X}\right)-\boldsymbol Z^{\top} \boldsymbol{\beta}_0\right\}^2 C|S=1\right]$, we have:
$$
\mathbb{E}\left[\boldsymbol Z \boldsymbol{C} \boldsymbol Z^{\top} \boldsymbol{\beta}_0-\boldsymbol Z \boldsymbol{C} \tilde{\eta}\left(\boldsymbol{X}\right)|S=1\right]=0.
$$
Furthermore, from the definition of $\boldsymbol{\beta}^{*}$, we have
$$
\tilde{\eta}\left(\boldsymbol{X}\right)=\boldsymbol Z^{\top} \boldsymbol{\beta}^{*}-b_a\left(\boldsymbol{X}\right)+o_p\left(K_n^{-(p+1)}\right).
$$
Hence, we obtain
$$
\mathbb{E}\left[\boldsymbol Z \boldsymbol{C} \boldsymbol Z^{\top}\left(\boldsymbol{\beta}^{*}-\boldsymbol{\beta}_0\right)-\boldsymbol Z \boldsymbol{C} b_a\left(\boldsymbol X\right)+\boldsymbol Z \boldsymbol{C} o_p\left(K_n^{-(p+1)}\right)|S=1\right]=0.
$$
The $k$-th compoment of first $\left(K_n+p\right)$ block of $\mathbb{E}\left[\boldsymbol Z\boldsymbol{C} b_a\left(\boldsymbol X\right)|S=1\right]$ is
$$
\begin{aligned}
& \mathbb{E}\left[\boldsymbol Z \boldsymbol{C} b_a\left(\boldsymbol{X}\right)|S=1\right]_k \\
= & \sum_{j=1}^D \int_{[0,1]^D} \frac{h^{\prime}\left(\tilde{\eta}\left(x\right)\right)^2}{h\left(\tilde{\eta}\left(x\right)\right)\left(1-h\left(\tilde{\eta}\left(x\right)\right)\right)} B_{-p+k}\left(x_{j}\right) b_{j, a}\left(x_{j}\right) d P\left(x\right)(1+o(1)) \\
= & o\left(K_n^{-(p+2)}\right).
\end{aligned}
$$
The last equality can be obtained by mimicking the proof of Proposition 3.1 in \cite{a31}.  It can be also derived that $\mathbb{E}\left\{\boldsymbol Z \boldsymbol{C}\right\}=O\left(K_n^{-1} \mathbf{1}\right)$. Hence we have
$$
\begin{aligned}
& \mathbb{E}\left[\boldsymbol Z \boldsymbol{C} \boldsymbol Z^{\top}\left(\boldsymbol{\beta}^{*}-\boldsymbol{\beta_0}\right)-\boldsymbol Z \boldsymbol{C} b_a\left(\boldsymbol{X}\right)+\boldsymbol Z \boldsymbol{C} o_p\left(K_n^{-(p+1)}\right)|S=1\right] \\
= & \mathbb{E}\left[\boldsymbol Z\boldsymbol{C} \boldsymbol Z^{\top}\left(\boldsymbol{\beta}^{*}-\boldsymbol{\beta}_0\right)|S=1\right]+o\left(K_n^{-(p+2)}\right)=0.
\end{aligned}
$$
Since $E\left[\boldsymbol Z \boldsymbol{C} \boldsymbol Z^{\top}|S=1\right]=O\left(K_n^{-1} \mathbf{1} \mathbf{1}^{\top}\right)$. We have
$$
\boldsymbol{\beta}_0-\boldsymbol{\beta}^{*}=o\left(K_n^{-(p+1)} \mathbf{1}\right).
$$

 \newpage
\section{B. Supplements for the Numerical Studies in Sections 3-5}
\setcounter{table}{0}   

\begin{figure}
\centering
\includegraphics[width = 13cm, height = 12cm]{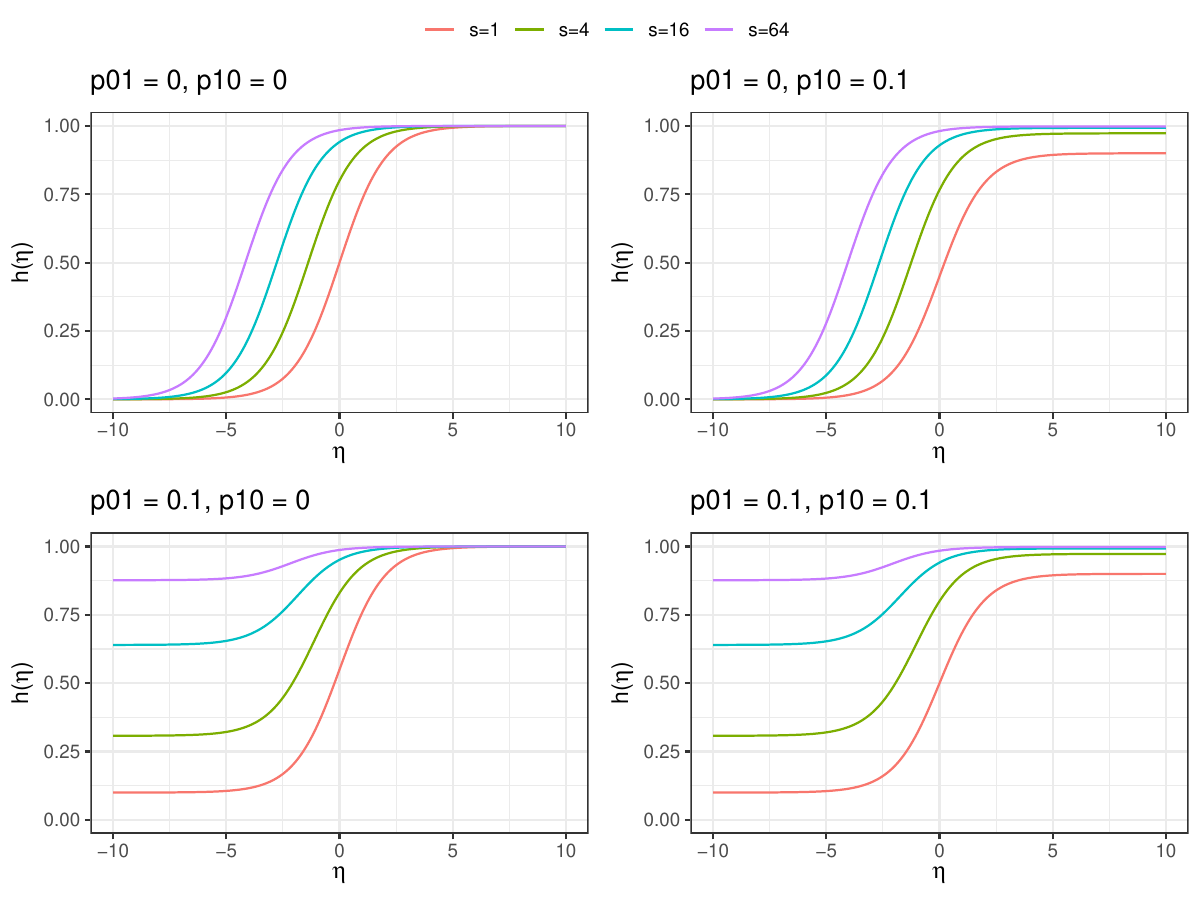}
\caption{The shape of adjusted link function $h(\eta)$ under different $p_{01}$,  $p_{10}$ and $s$.}
\label{f:figure2}
\end{figure}

\begin{figure}
\centering
\includegraphics[width = 12cm, height = 6cm]{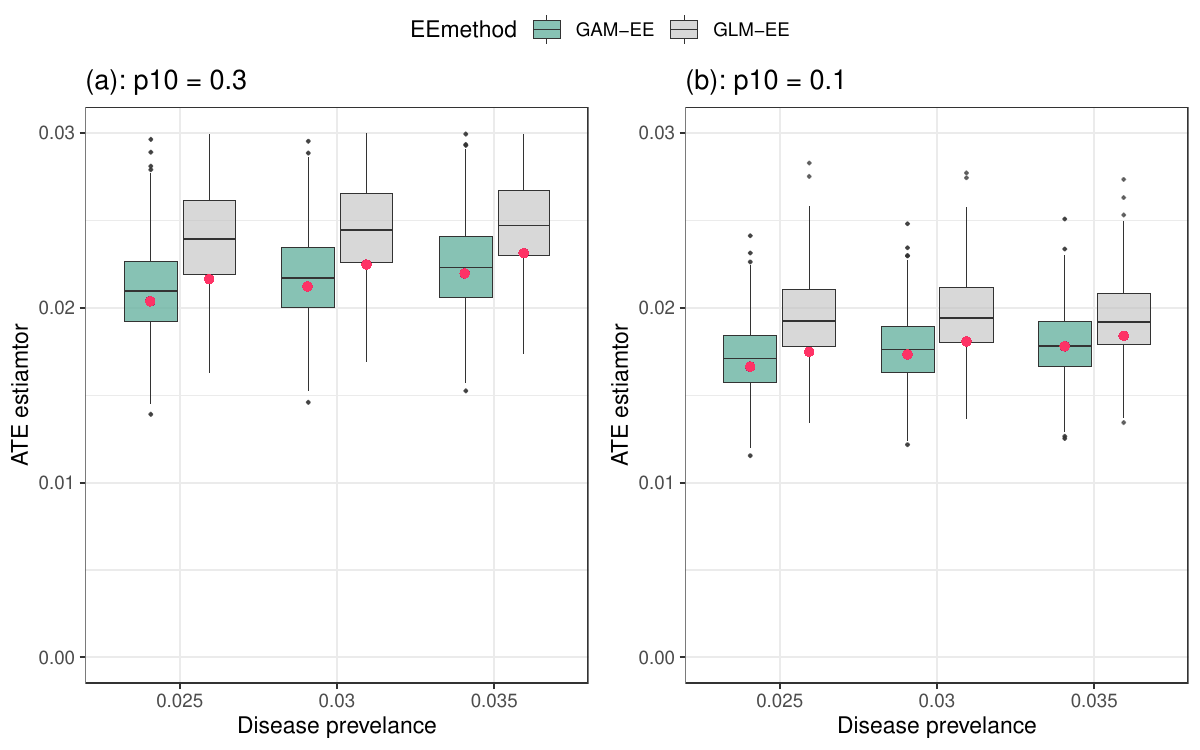}
\caption{Estimators of $\tau$ under different predetermined disease prevalence and false negative rates. The red solid points are the benchmarks of the standard EE methods applied in the full dataset from UK biobank.}
\label{f:figure5.2}
\end{figure}

\setcounter{table}{0}   

\begin{table}
\caption{Summary of simulation results for $p_{01}>0$.}
\label{t:table s1}
\begin{center}
\resizebox{\textwidth}{!}{
\begin{tabular}{lllllllllllllll}
\hline
\midrule
       &    &    & \multicolumn{6}{c}{$p_{01}=0.03$}                           & \multicolumn{6}{c}{$p_{01}=0.06$}                          \\ \cmidrule(lr){4-9} \cmidrule(lr){10-15}  
       &    &    & \multicolumn{3}{c}{GAM-EE} & \multicolumn{3}{c}{GLM-EE} & \multicolumn{3}{c}{GAM-EE} & \multicolumn{3}{c}{GLM-EE} \\ \cmidrule(lr){4-6} \cmidrule(lr){7-9} \cmidrule(lr){10-12} \cmidrule(lr){13-15} 
       model &    $p_{10}$&     $v$  & Rbias   & RMSE   & CP    & Rbias   & RMSE   & CP    & Rbias   & RMSE   & CP    & Rbias   & RMSE   & CP    \\ \midrule
      
\textbf{M1}&$0$ & 0.05 & 0.25     & 0.538   & 95.4  & 0.24     & 0.538   & 95.2  & -0.85    & 0.647   & 95.2  & -5.59    & 0.833   & 87.4  \\
&       & 0.1  & 0.42     & 0.781   & 96.4  & 0.50     & 0.781   & 95.6  & -0.74     & 0.869   & 94.8  & -0.49     & 0.913  & 95.4  \\ 

&$0.2$  & 0.05 & 0.25     & 0.586   & 95.6  & 0.23     & 0.593   & 95.4  & -0.64     & 0.76   & 94.6  & -6.64     & 0.994   & 85.0  \\
&       & 0.1  & 0.25     & 0.868   & 94.4  & 0.39     & 0.875   & 94.2  & -1.39     & 0.982   & 93.2  & -1.28     & 1.109   & 94.2  \\
       
&$0.4$  & 0.05 & 0.40     & 0.639   & 95.8  & 0.32     & 0.642   & 95.8  & -1.28     & 0.889   & 94.2  & -9.23     & 1.248   & 85.8  \\
&       & 0.1  & 0.17     & 0.964   & 95.6  & 0.41     & 0.973   & 95.8  & -2.61     & 1.231   & 93.0  & -3.00     & 1.319   & 92.6  \\
\midrule
\textbf{M2}&$0$ & 0.05 & 0.10 &	0.519 &	95.2 &	-1.21 &	0.540 &	94.8& 0.11 &	0.673 &	94.2 &	-5.95 &	0.865 &	86.0\\
&       & 0.1  & -0.35 &	0.703 &	96.8 &	1.80 &	0.794 &	95.2 & -0.20 &	0.927 &	93.2 &	-0.64 &	0.952 &	93.8\\ 
&$0.2$  & 0.05 & 0.23 &	0.596 &	94.8 &	-2.18 &	0.631 &	92.6 & 0.85	& 0.793	& 92.0 &	-6.85 & 1.017 & 84.2\\
&       & 0.1  & -0.64 &	0.833 &	94.2 &	1.36 & 	0.924 &	93.8  & -0.03	 & 0.974 &	95.4 &	-0.95 &	1.031 &	94.2\\
&$0.4$  & 0.05 & 0.21 &	0.681 &	95.4 &	-3.79	& 0.760 &	92.2 & 2.34 &	0.959 &	93.5 &	-8.15 &	1.198 &	86.2\\
&       & 0.1  & -0.73 &	0.961 &	94.4 & 0.82 &	1.011 &	94.6 & -0.36 &	1.199 & 95.2 &	-2.07 &	1.277 &	92.4\\
\midrule
\textbf{M3}&$0$ & 0.05 &-0.19	&	0.538	&	94.4	&	2.35	&	0.600	&	93.2	&	2.28	&	0.684	&	93.8	&	4.59	&	0.792	&	90.2\\
&       & 0.1  &-0.38	&	0.741	&	95.8	&	-0.27	&	0.751	&	96.0	&	0.40	&	0.891	&	94.0	&	1.27	&	0.912	&	93.8\\ 
&  $0.2$  & 0.05 & 0.05	&	0.561	&	95.2	&	3.46	&	0.671	&	92.6	&	1.84	&	0.725	&	95.0	&	4.77	&	0.870	&	92.2\\
&       & 0.1  &-0.48	&	0.864	&	93.6	&	0.99	&	0.888	&	94.2	&	0.76	&	1.013	&	95.2	&	3.20	&	1.177	&	91.2\\
&$0.4$  & 0.05 & 1.01	&	0.642	&	95.0	&	5.05	&	0.823	&	90.2	&	2.75	&	0.911	&	93.8	&	5.53	&	1.061	&	91.6\\
&       & 0.1  &-0.89	&	0.975	&	94.0	&	1.78	&	1.023	&	94.8	&	0.69	&	1.102	&	95.6	&	4.17	&	1.345	&	91.8\\
\midrule
\textbf{M4}&$0$ & 0.05 &-1.69	&	0.522	&	94.4	&	0.84	&	0.524	&	95.4	&	0.26	&	0.615	&	94.0	&	-0.15	&	0.646	&	93.2 \\
&       & 0.1  &-1.14	&	0.760	&	93.4	&	1.14	&	0.817	&	94.4	&	-0.06	&	0.830	&	95.8	&	1.17	&	0.874	&	95.2\\ 
&$0.2$  & 0.05 &-2.27	&	0.573	&	92.4	&	0.69	&	0.598	&	95.4	&	-1.19	&	0.707	&	94.8	&	-1.50	&	0.767	&	93.8\\
&       & 0.1  &-2.73	&	0.862	&	90.4	&	0.82	&	0.834	&	95.8	&	-1.21	&	0.939	&	95.0	&	1.16	&	0.994	&	94.8\\
&$0.4$  & 0.05 & -1.88	&	0.634	&	94.4	&	1.15	&	0.676	&	95.2	&	0.58	&	0.790	&	95.8	&	-0.39	&	0.864 & 96.4 \\
&       & 0.1  &-2.49	&	0.942	&	92.6	&	2.02	&	0.958	&	95.2	&	-1.37	&	1.075	&	95.2	&	1.65	&	1.165	&	96.4 \\
\midrule  
\hline
\end{tabular}
}
\begin{tablenotes}
        \footnotesize
        \item[1] Rbias, relative bias (\%); RMSE, root mean squared error ($\times$ 1000); CP, coverage probability (\%).
\end{tablenotes}
\end{center}
\end{table}

\begin{table}[]
\label{t:table s2}
\caption{Demographic and lifestyle characteristics of 136741 persons by Alcohol Intake Category in the UK Biobank Study}
\begin{tabular}{lllll}
\hline
\midrule
                              & Overall       & Non-consumer            & Consumer            & p-value                \\ \midrule
n                             & 136,741       & 63,478       & 73,263       &                  \\
gout = 1 (\%)                 & 5264 (3.8)    & 2010 (3.2)   & 3254 (4.4)   & \textless{}0.001 \\
edu = 1 (\%)                  & 49242 (36.0)  & 19530 (30.8) & 29712 (40.6) & \textless{}0.001 \\
ethnicity = 1 (\%)            & 125352 (91.7) & 57022 (89.8) & 68330 (93.3) & \textless{}0.001 \\
TDI = 1 (\%)             & 68370 (50.0)  & 34189 (53.9) & 34181 (46.7) & \textless{}0.001 \\
exercise = 1 (\%)             & 68345 (50.0)  & 31798 (50.1) & 36547 (49.9) & 0.446            \\
diet score (mean (SD))       & -0.61 (1.03)  & -0.56 (1.07) & -0.65 (1.00) & \textless{}0.001 \\
age (mean (SD))               & 57.13 (8.06)  & 56.37 (8.37) & 57.78 (7.72)  & \textless{}0.001 \\
household income (mean (SD)) & 2.69 (1.19)   & 2.49 (1.16) & 2.86 (1.18)  & \textless{}0.001 \\
BMI (mean (SD))               & 27.84 (4.14)   & 28.20 (4.45)  & 27.52 (3.84)  & \textless{}0.001
 \\ \midrule \hline
\end{tabular}
\begin{tablenotes}
        \footnotesize
        \item[1] gout (coded as 1 for diagnosed with gout);
edu (coded as 1 for college education and 0 for others); ethnicity (coded as 1 for White ethnicity and 0 for others); TDI (Discrete Townsend Deprivation Index, with TDI coded as 1 for the top 50\% individuals and 0 for others); exercise (coded as 1 for individuals in the top 50\% of daily exercise time and 0 for others); diet score (a measure of healthy diet adherence); household income (classified into ordered discrete categories: 1 for less than $18,000$, 2 for $18,000$ to $30,999$, 3 for $31,000$ to $51,999$, 4 for $52,000$ to $100,000$, and 5 for greater than $100,000$); BMI (calculated as weight divided by height squared);
\end{tablenotes}
\end{table}

\label{lastpage}

\end{document}